\documentclass[a4paper,11pt]{article}
\usepackage{jcappub} % for details on the use of the package, please see the JINST-author-manual
\usepackage{dsfont} % for \mathds{R}, \mathds{C}, etc
\usepackage{mathtools} % for \coloneqq
\usepackage{amsfonts}

\usepackage{tikz} % for k configuration diagrams

\allowdisplaybreaks

\newcommand{\N}{\ensuremath{\mathbb{N}}}	        % Natural numbers N
\newcommand{\R}{\ensuremath{\mathbb{R}}}	        % Real numbers R
\newcommand{\C}{\ensuremath{\mathbb{C}}}          % Complex numbers C

\newcommand{\e}{\mathrm{e}}					        % Euler number e
\newcommand{\iu}{\mathrm{i}}				        % Imaginary unit i
\newcommand{\idmat}{\ensuremath{\mathds{1}}}	    % Identity matrix

\newcommand{\mathbfit}[1]{\textbf{\textit{#1}}}
\newcommand{\mathbfss}[1]{\textbf{\textsf{#1}}}
\newcommand{\myvec}[1]{\mathbfit{#1}}               
\newcommand{\mymat}[1]{\mathbfss{#1}}               

\newcommand{\abs}[1]{\ensuremath{\lvert#1\rvert}}   % Absolute value of #1

\newcommand{\bispec}{\mathcal{B}}    % Symbol for the bispectrum
\newcommand{\powspec}{\mathcal{P}}    % Symbol for the power spectrum
\newcommand{\diracd}[1]{\delta_{\text{D}}\!\left(#1\right)} % Dirac delta function
\newcommand{\expr}[1]{\e^{#1}}                      % Exponential function
\newcommand{\corr}[1]{\left\langle #1\right\rangle}   % Correlation function
\newcommand{\dpart}[1]{\frac{\partial}{\partial\:\! #1}}    % Partial derivative
\newcommand{\dpartsq}[2]{\frac{\partial^2}{\partial\:\! #1\:\!\:\!\partial\:\! #2}}    % Second partial derivative
\newcommand{\Zel}{Zel'dovich}

\arxivnumber{2504.17386}
\title{Free Cosmic Density Bispectrum on Small Scales}

\author{Ricardo Waibel,}
\author{Sara Konrad}
\author{and Matthias Bartelmann}
\affiliation{Institute for Theoretical Physics, Heidelberg University, Philosophenweg 12 and 16, 69120 Heidelberg, Germany}

\emailAdd{waibel@thphys.uni-heidelberg.de}

\abstract{We study the asymptotic behaviour of the free, cold-dark matter density fluctuation bispectrum in the limit of small scales. From an initially Gaussian random field, we draw phase-space positions of test particles which then propagate along \Zel\ trajectories. Only initial momentum-momentum correlation are considered, making the formulas identical to the typical Zel'dovich approximation. A suitable expansion of the initial momentum auto-correlations of these particles leads to an asymptotic series whose lower-order power-law exponents we calculate. The dominant contribution has an exponent of $-11/2$. For triangle configurations with zero surface area, this exponent is even enhanced to $-9/2$. These power laws can only be revealed by a non-perturbative calculation with respect to the initial power spectrum. They are valid for a general class of initial power spectra with a cut-off function, required to enforce convergence of its moments. We then confirm our analytic results numerically. Finally, we use this asymptotic behaviour to investigate the shape dependence of the bispectrum in the small-scale limit, and to show how different shapes grow over cosmic time. These confirm the usual model of gravitational collapse within the \Zel\ picture.}

\begin{document}
	\maketitle
	\flushbottom
	
	\section{Introduction}
	
	Identifying and understanding the origin and evolution of generic properties of cosmic structures on small, non-linear scales is an important aspect of cosmology. At the time of the CMB release, typical amplitudes of relative density fluctuations are $\sim 10^{-5}$, reflected by the CMB temperature fluctuations \cite{akra18}. In this initially nearly-homogeneous universe, the initial density field can be modelled as a Gaussian random field. It is then fully characterized by its two-point correlation function or equivalently its power spectrum. For the evolution of this initial state, gravity is most important, especially on large scales. Relative to the expanding background of an underlying Friedmann-Lemaître-Robertson-Walker spacetime, Newtonian gravitational forces suffice to describe the local dynamics of dark matter and baryons. 
	
	Several approaches exist to address the evolution of cosmic structures. Numerical simulations allow for extensive tests of many effects relevant for the full picture of interactions and feedback processes \cite{springel2012larger, vogelsberger2020cosmological}. However, they are computationally expensive and thus only available for a limited set of cosmological parameters. Analytical approaches tend to need more idealizing assumptions than simulations, but can be used to scan wide parameter ranges and theory classes, and can provide insight into fundamental reasons why the universe appears as we observe it today.
	
	Most of the analytic approaches build on the matter density and the velocity as fundamental fields. This is problematic in regions where multiple streams are expected to form due to the collision-less nature of the dark matter, which is assumed to be the main constituent of cosmic structures. This limits the applicability of perturbative approaches based on smooth fields, such as Eulerian standard or Lagrangian perturbation theory \cite{ma1995cosmological, bern02, buchert1992lagrangian}. To by-pass this notorious shell-crossing problem, kinetic field theory (KFT) dissolves cosmic structures into a statistical, non-equilibrium ensemble of classical particles whose trajectories are traced through phase space \cite{bart16, bart19}. Since Hamiltonian trajectories cannot cross in phase space, KFT can enter into small, non-linear scales without conceptual limitations.
	
	The central quantity of many structure-formation studies is the power spectrum, i.e., the mean-squared amplitude of structures as a function of their scale. For Gaussian random fields, this information is complete. Initially Gaussian random fields however become more complex as they evolve and build up higher-order spectra. The first one beyond the power spectrum is the bispectrum, defined as the Fourier transform of the three-point function. It adds information on the abundance of structures of a given shape \cite{lewi11}, and it allows to quantify how strongly non-Gaussian processes affect the evolution of cosmic structures. This is very valuable information which is sensitive to modifications of the underlying physics, and which will become fully observationally accessible within the current stage IV cosmological surveys. 
	
	The bispectrum has conventionally been calculated from Eulerian perturbation theory \cite{fry1984galaxy, bern02}. The results can be understood as an approximation valid on large scales. They rely on the linearized system of the continuity, Euler, and Poisson equations. On small scales, however, this approach ceases to be valid. Alternatively, Lagrangian perturbation theory models structure formation based on the displacement of fluid elements from an initial position. Recently, the approach was used to calculate the \Zel\ bispectrum numerically \cite{Chen_2024}, focussing on mildly non-linear scales.
	
	In this work, we derive the asymptotic behaviour of the bispectrum on small scales, based on kinetic field theory. This is done to the lowest order for the full expression, which is the free theory based on the \Zel\ approximation \cite{zel1970gravitational, bart16}. Since its conception, this approximation has been applied as a first approximation to the evolution of cosmic structures. It describes particle trajectories as straight lines in a time coordinate which is a non-linear function of cosmic time, starting from initially correlated positions and momenta. We will only take momentum correlations into account because they dominate the evolution at moderate and late cosmic times \cite{bart16}. Doing this, the description is then equivalent to the Zel'dovich approximation known from the context of Eulerian and Lagrangian Perturbation Theory.
	
	The asymptotic behaviour of the bispectrum is important for studying the formation of small-scale structures like halos, and the gravitational collapse leading to them. It also provides interesting insight into virialization, understood as the process of turning ordered into unordered kinetic energy. The bispectrum has also been used for analyzing redshift distortions, with a focus on moderate scales \cite{hivon1994redshift}.
	
	While our results to be derived below are based on the \Zel\ approximation and thus neglect small-scale gravitational interactions, they provide an important step into the direction of understanding rigorous and inevitable aspects of small-scale cosmic structures. As we will show, starting with fairly general assumptions on the initial state, the non-perturbative approach with respect to the initial correlations reveals a richer dependence at leading order on the initial power spectrum than expected from the large-scale or even previous small-scale approximations to the power spectrum \cite{konrad2022kinetic}. More precisely, the bispectrum can be approximated as a square of power spectra \cite{fry1982transform, fry1984galaxy, bern02} in a large-scale approximation, which implies, to leading order, a power-law scaling $\propto k^{-6}$ for large wave numbers $k$, as this is expected from the small-scale behaviour of the power spectrum, even beyond the \Zel\ approximation \cite{konrad2022kinetic, ginat2025gravitational, chen2020asymptotic}.
	
	We structure this paper as follows: Section~\ref{sec:methods} will collect all the methods needed to derive the asymptotic behaviour of the bispectrum for small scales from KFT, going into different special cases that need to be considered. It also describes how the free KFT expression can be evaluated numerically, providing means to validate the derived results. Section~\ref{sec:results} contains a comparison of the full numerical evaluation of the free bispectrum and of the large- and small-scale approximations to it. Also, physical insight from the resulting formulae is discussed. Section~\ref{sec:summary} summarizes our results and ends with an outlook. The Appendix collects detail on the calculations performed in Sect.~\ref{sec:methods}.
	
	We conclude this introduction with a brief remark on notation. Vectors are denoted as $\myvec{v}$, matrices as $\mymat{M}$. Vector components are labelled as $\myvec{v}=(v_x,v_y,x_z)^\top$. When clear from context, a plain vector $\myvec{q}$ will be used to refer to the six-dimensional collection of two three-dimensional vectors $\myvec{q}_2$ and $\myvec{q}_3$. Often, the Euclidean norm of a vector $\myvec{k}$ will simply be denoted by $k$. We will use the following short-hand notation for integrals,
	\begin{equation}
		\int_{\myvec{q}} = \int \mathrm{d}^3\myvec{q}\;,\quad
		\int_{\myvec{k}} = \int \frac{\mathrm{d}^3\myvec{k}}{(2\pi)^3}\;,
	\end{equation}
	and the Fourier convention is fixed by 
	\begin{equation}
		\tilde{f}(\myvec{k}) = \int_{\myvec{q}} f(\myvec{q})
		\expr{-\iu \myvec{k}\cdot \myvec{q}} \;,\quad
		f(\myvec{q}) = \int_{\myvec{k}} \tilde{f}(\myvec{k})
		\expr{\iu \myvec{k}\cdot \myvec{q}}\ .
	\end{equation}
	
	\section{Methods} \label{sec:methods}
	
	\subsection{Free Bispectrum from Kinetic Field Theory}
	
	We use kinetic field theory (KFT) to describe cosmic structure formation. For details of its formalism, we refer the reader to the review \cite{konrad2022kinetic} which deals specifically with the free theory, neglecting particle interactions beyond the \Zel\ approximation.
	
	The bispectrum $\bispec$ is the connected, synchronous, 3-point correlation function of the density contrast, which can be calculated in KFT from a generating functional $Z$,
	\begin{equation}
		(2\pi)^3 \diracd{\sum_{i=1}^{3} \myvec{k}_i} \bispec(\myvec{k}_2,\myvec{k}_3) = \corr{\delta_1(\myvec{k}_1)\delta_2(\myvec{k}_2)\delta_3(\myvec{k}_3)}_c \approx \frac{N^3}{\bar{\rho}^3} Z[\myvec{L}]\ , \label{eq:bispec:generalderivation}
	\end{equation}
	with $\delta$ the density contrast, $N$ the total particle number, $\bar\rho$ the mean dark-matter density, and $\myvec{L}$ a shift vector resulting from applying three density operators to the full generating functional. We drop the time argument from $\bispec$ for brevity. The linear growth factor $D_+$ is used as the time coordinate here for convenience, $t = D_+-1$, setting $t=0$ by normalising $D_+$ to unity at the initial cosmic time.
	
	Calculating the free generating functional $Z_0[\myvec{L}]$ from KFT using only momentum-momentum correlations, the expressions are identical to the Zel'dovich approximation bispectrum known from Eulerian and Lagrangian perturbation theory. The final equation for the free bispectrum is
	\begin{equation}
		\diracd{\sum_{i=1}^{3} \myvec{k}_i} \bispec(\myvec{k}_2,\myvec{k}_3) = \diracd{\sum_{i=1}^{3} \myvec{k}_i} \expr{-Q_D} \int_{\myvec{q}_{12}}\int_{\myvec{q}_{13}} \exp\left(-t^2 Q_C\right) \expr{\iu(\myvec{k}_2\cdot \myvec{q}_{12}+\myvec{k}_3\cdot \myvec{q}_{13})} \label{eq:bispec:general}
	\end{equation}
	with the Dirac distribution $\delta_\mathrm{D}$ ensuring the triangle condition for the $\myvec{k}$ vectors. The exponent functions are
	\begin{align}
		Q_D &= \frac{\sigma_1^2t^2}{6}\sum_{i=1}^{3} \myvec{k}_i^2 \quad\mbox{and} \label{eq:bispec:quadform1}\\
		Q_C &= \myvec{k}_1^t \mymat{C}_{p_1p_2}(\myvec{q}_{12})\myvec{k}_2 + \myvec{k}_1^t \mymat{C}_{p_1p_3}(\myvec{q}_{13})\myvec{k}_3 +\myvec{k}_2^t \mymat{C}_{p_2p_3}(\myvec{q}_{12}-\myvec{q}_{13})\myvec{k}_3\ . \label{eq:bispec:quadform2}
	\end{align}
	Here, $\mymat{C}_{pp}$ is the auto-correlation matrix of the initial particle momenta. Since these dominate the amplitude of cosmic structures by orders of magnitude at moderate and late cosmic times, it is well justified to ignore density-density and density-momentum correlations \cite{bart16}. The momentum auto-correlation matrix is determined by the power spectrum $P_\psi$ of the initial velocity potential by
	\begin{equation}
		\mymat{C}_{p_i p_j}(\myvec{q}) = \int_{\myvec{k}} \left(\myvec{k} \otimes \myvec{k}\right) P_\psi(k) \expr{\iu \myvec{k}\cdot\myvec{q}} = \idmat_3 a_1(q) - (\hat{q} \otimes \hat{q})\, a_2(q) \ . \label{eq:cpp:definition}
	\end{equation}
	The correlation functions $a_1$ and $a_2$ introduced here will turn out to be decisive. The function $a_1$ is the momentum-autocorrelation function perpendicular to the line connecting the two momenta, while $a_1-a_2$ is the momentum-autocorrelation function parallel to it. They can be written as
	\begin{align}
		a_1(q) &= -\frac{1}{2\pi^2} \int_0^\infty \mathrm{d} k\,
		P_\delta^\mathrm{(i)}(k)\, \frac{j_1(k q)}{k q}\ , \label{eq:bispec:a1}\\
		a_2(q) &= \frac{1}{2\pi^2} \int_0^\infty \mathrm{d} k\,
		P_\delta^\mathrm{(i)}(k)\, j_2(k q)\ , \label{eq:bispec:a2}
	\end{align}
	with $P_\delta^\mathrm{(i)}$ the initial density contrast power spectrum. Two further comments are in order on the general form (\ref{eq:bispec:general}) of the free bispectrum. Firstly, the integrand without the complex phase is an even function with respect to a joint reflection of the $\myvec{q}$ vectors. The bispectrum is thus real-valued, which means that the complex phase can be replaced by a cosine function. Secondly, due to isotropy, the $\myvec{k}$ vectors can be chosen to point into a direction convenient for numerical evaluation. We will do so when needed, and then use the convention $\myvec{k}_2 = k_{2z}\hat{e}_z$ and $\myvec{k}_3 = k_{3x}\hat{\myvec{e}}_x + k_{3z}\hat{\myvec{e}}_z$.
	
	Finally, there is an approximation for large scales, i.e., the limit $k\rightarrow 0$, which has been previously derived \cite{bart17}. The small $k$ justifies an expansion around zero of the exponential function containing $Q_C$ in (\ref{eq:bispec:general}) up to second order. This allows to fully integrate the expression, leading to the approximation
	\begin{equation}
		\bispec(\myvec{k}_2,\myvec{k}_3) \approx
		F(\myvec{k}_2,\myvec{k}_3) P^{(\text{lin})}(k_2) P^{(\text{lin})}(k_3) +
		\text{cyc.}
		\label{eq:bispec:largescale}
	\end{equation}
	with $P^{(\text{lin})}$ the linearly evolved initial power spectrum and the kernel function
	\begin{equation}
		F(\myvec{k}_2,\myvec{k}_3) = \left(1+\frac{\myvec{k}_2 \cdot \myvec{k}_3}{k_2^2}\right) \left(1+\frac{\myvec{k}_2 \cdot \myvec{k}_3}{k_3^2}\right)\ .
	\end{equation}
	This is structurally similar to the result from Eulerian perturbation theory \cite{bern02}. The difference reflects a different way of including self-gravity \cite{bart16}.
	
	In view of this result, it is natural to define the reduced bispectrum as
	\begin{align}
		Q = \frac{\bispec(\myvec{k}_2,\myvec{k}_3)}
		{\powspec(k_1)\powspec(k_2)+
			\powspec(k_2)\powspec(k_3)+
			\powspec(k_3)\powspec(k_1)} \ ,
	\end{align}
	where $\powspec(k)$ is the power spectrum. This was first introduced in \cite{fry1982transform}. As we often approximate $\bispec$, we need to approximate $\powspec$ as well. To remain consistent, we will approximate it by the linearly evolved initial power spectrum $\powspec(k)\approx P^{(\text{lin})}(k)$. This is then also in line with the approximation in Eq.~(\ref{eq:bispec:largescale}). In many scenarios, considering $Q$ instead of $\bispec$ effectively removes most of the cosmology dependence, especially in the large-scale limit \cite{bern02}.
	
	\subsection{Small-Scale Approximation of the Free Bispectrum}
	
	Aiming at a small-scale description of the free bispectrum in Eq.~(\ref{eq:bispec:general}), an approach based on Laplace-type integrals is possible, used for example in \cite{konr20, 2022MNRAS.515.2578K, 2022MNRAS.515.5823K} for the free power spectrum. The central argument there rests on the Morse lemma, which allows to approximate the relevant exponent. This is not possible here, as the Hessian of the exponent is singular. Thus we will apply a method based on the splitting lemma of functions as a special case of a saddle point approximation with degenerate critical points. Since the lemma exists in several variants \cite{broc75, post78, wong01}, we will focus on the presentation in terms of power series as in \cite{greu07}. 
	
	Let $\C\{x_1,\dots,x_n\}$ be the ring of convergent power series with the standard norm, and let $\operatorname{ord}(f)$ be the order of the power series $f\in \C\{x_1,\dots,x_n\}$, i.e.\ the lowest sum of exponents of monomials with non-zero coefficients. Let $\mymat{H}(f)(\myvec{x})$ denote the Hessian of $f$ at $\myvec{x}$, where $\myvec{x}$ is the collection of all variables $x_1,\dots,x_n$.
	
	Then, the splitting lemma states that, for $f\in\C\{x_1,\dots,x_n\}$ with $\operatorname{ord}(f)\geq 2$ and $\operatorname{rank}(\mymat{H}(f)(0))=k$, $f$ can be represented as
	\begin{equation}
		f \sim x_1^2 + \dots + x_k^2 + g(x_{k+1},\dots,x_n)\ ,
		\label{eq:splittinglemma:general}
	\end{equation}
	with $g\in\C\{x_{k+1},\dots,x_n\}$, and $\operatorname{ord}(g)\geq 3$ or $g=0$. This $g$ is uniquely determined up to right equivalence, where right equivalence and the $\sim$ symbol refer to local biholomorphic coordinate transformations.
	
	For a proof of this lemma, see e.g.\ \cite{greu07}. It relies on the critical point at zero being isolated, which then implies that the series is finitely determined. This is defined as follows: for a finitely determined power series $f$, there exists a degree $n$ such that for all power series $g$ with terms identical to $f$ up to degree $n$, this immediately implies $f\sim g$. Also, the proof has been extended to the real domain, with the only difference that signs have to be considered in front of the monomials \cite{mara15}.
	
	In this form, however, the lemma does not yet help with the integral we wish to calculate. Nonetheless, its constructive proof can be followed in parts to transform the integral into a helpful representation. The forms defined in Eqs.~(\ref{eq:bispec:quadform1}) and (\ref{eq:bispec:quadform2}) add up to
	\begin{equation}
		Q_f = Q_D + t^2 Q_C\ ,
		\label{eq:bispec:quadformfull}
	\end{equation}
	which has a critical point at zero and also evaluates to zero at $\myvec{q}=0$, using $\myvec{q}=(q_1,\dots,q_6)$ as the 6-dimensional collection of $\myvec{q}_{12}$ and $\myvec{q}_{13}$. The calculation is carried out in Appendix~\ref{sec:app:checksplit}. The Hessian at $\myvec{q}=0$ is given by (using isotropy to reduce the dependence on the full $\myvec{k}_2$, $\myvec{k}_3$ to 3 components)
	\begin{equation}
		\mymat{H} = \frac{t^2\sigma_2^2}{15} \begin{pmatrix} \mymat{H}_{22} && \mymat{H}_{23}\\
			\mymat{H}_{32} && \mymat{H}_{33}
		\end{pmatrix}
	\end{equation}
	with the component matrices
	\begin{equation}
		\mymat{H}_{ab} = \myvec{k}_a \otimes \myvec{k}_b +
		\myvec{k}_b \otimes \myvec{k}_a +
		\idmat_3\left(\myvec{k}_a\cdot \myvec{k}_b\right)\ ,\  a,b\in\{2,3\}\ .
	\end{equation}
	This Hessian $\mymat{H}$ has a rank of $5$ out of a possible $6$, hence corank $1$, if $k_{3x}\neq 0$. This makes it particularly easy because the splitting lemma then implies that five variables can be separated into a Gaussian in the integral and only one variable remains in a more complicated expression. For $k_{3x}=0$, the corank increases to $3$, but this corresponds to a configuration with parallel $\myvec{k}_2$ and $\myvec{k}_3$, hence to a degenerate configuration, as the spanned triangle has zero surface area. This will be treated separately in Sect.~\ref{sec:methods:parallel}.
	
	The key condition of the splitting lemma, i.e., that the critical point be isolated, is also shown in Appendix~\ref{sec:app:checksplit}. This is in general quite difficult to establish, but can be calculated explicitly in the non-degenerate case. 
	
\begin{figure*}
		\includegraphics[width=\textwidth]{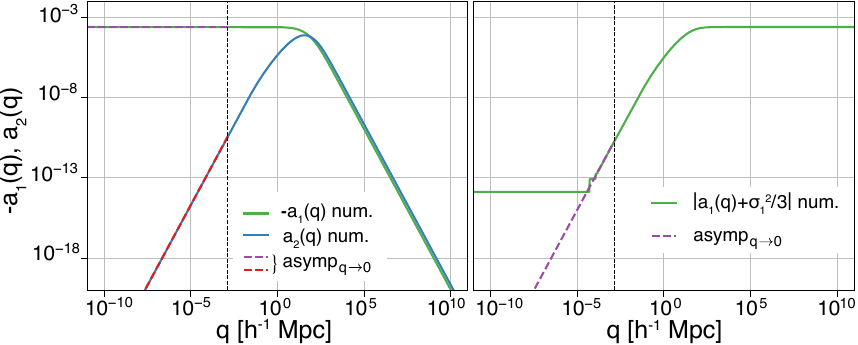}
		\caption{The left panel shows the correlation functions $a_1$ and $a_2$ (cf.\ Eqs.~(\ref{eq:bispec:a1}),~(\ref{eq:bispec:a2})). The vertical dashed line indicates the heuristic value up to which point the asymptotic description (cf.\ Eqs.~(\ref{eq:a1:definition})-(\ref{eq:a2:definition})) is used for the functions instead of their full integral expressions. In the right panel, the important combination $a_1+\sigma_1^2/3$ is shown as a naive addition and using the asymptotic expansion to provide numerical stability for small $q$. As later computations require this stability for very small $q$, an implementation of these functions using asymptotic descriptions is important. \\
		The initial power spectrum is taken as the Bardeen power spectrum \cite{bard86} using an exponential cut-off with $k_s=10^2 \, h\operatorname{Mpc}^{-1}$.}
		\label{fig:integrand:a1a2}
	\end{figure*}

	The last prerequisite is then to rewrite $Q_f$ from (\ref{eq:bispec:quadformfull}) as a convergent power series in the neighbourhood of the critical point $\myvec{q}=0$. We adopt from \cite{konr20, 2022MNRAS.515.2578K} the power series
	\begin{align}
		a_1(q) &\approx -\frac{\sigma_1^2}{3} - \sum_{n=1}^\infty \frac{\sigma_{n+1}^2 (-q^2)^n}{(2n+3)(2n+1)!}\ , \label{eq:a1:definition}\\
		a_2(q) &\approx q^2 \sum_{n=0}^\infty \frac{\sigma_{n+2}^2 (-q^2)^n}{(2n+5)(2n+3)(2n+1)!}\ , \label{eq:a2:definition}
	\end{align}
	of the momentum-correlation functions $a_1$, $a_2$ for $q\rightarrow 0$, with the moments for $n\in\N$
	\begin{equation}
		\sigma_n^2 = \frac{1}{2\pi^2} \int_0^\infty \mathrm{d}k \, k^{2n-2}P^\mathrm{(i)}_\delta (k)\ .
	\end{equation}
	The asymptotic expressions are derived via an expansion of the spherical Bessel functions for small values. The functions can be seen in Fig.~\ref{fig:integrand:a1a2}, specifically the combination $a_1(q)+\sigma_1^2/3$ which will be important for numerical stability.
	
	It is also shown in \cite{konr20, 2022MNRAS.515.2578K} that the power series expressions for $a_1$, $a_2$ are convergent for the case that the initial power spectrum has an exponential cut-off. This cut-off is introduced to keep the moments $\sigma_n^2$ finite for all $n\in\N$. We will use an exponential cut-off of the following form
	\begin{equation}
		P^\mathrm{(i)}_\delta \rightarrow P^\mathrm{(i)}_\delta \expr{-\frac{k}{k_s}}\ ,
		\label{eq:spectral:cutoff}
	\end{equation}
	with which we can then rewrite $Q_f$ as a convergent power series in a neighbourhood of $\myvec{q}=0$. Introducing such a cut-off means that initially there is no structure smaller than a typical length scale of $k_s^{-1}$. Such a behaviour can for example be explained by dark matter that is not completely cold. Calculations for a purely cold dark matter case are possible and have been successfully applied to the Zel'dovich power spectrum in \cite{2022MNRAS.515.5823K}. However, the leading order asymptotics was not changed by this. Changing the scale $k_s$ will impact the concrete values of the $\sigma_n^2$ which in turn dictate the range of applicability of the asymptotic expressions.
	
	The first transformation in the proof of the splitting lemma rewrites the quadratic part of the power series in the desired form. Applying Sylvester's law of inertia, a basis can be found such that $\mymat{H}$ is a diagonal matrix with the only entries $-1$, $1$, and $0$. Relaxing this condition slightly such that the columns of the transformation matrix $\mymat{T}$ become simpler and $\mymat{T}$ has $\det(\mymat{T})=1$, the result is
	\begin{equation}
		\mymat{T} = \begin{pmatrix} 1&0&0&-\frac{k_{3z}}{k_{2z}}&0&-\frac{k_{3x}}{k_{2z}}\\
			0&1&0&0&-\frac{k_{3z}}{k_{2z}}&0\\
			0&0&1&-\frac{k_{3x}}{3k_{2z}}&0&-\frac{k_{3z}}{k_{2z}}\\
			0&0&0&1&0&0\\
			0&0&0&0&1&0\\
			0&0&0&0&0&1 \end{pmatrix}\ .
	\end{equation}
	This matrix is dimension-less because it contains only ratios of physical quantities. The Hessian $\mymat{H}$ after transformation is then
	\begin{equation}
		\frac{1}{2}\left(
		\mymat{T}^\top\mymat{H}\mymat{T}
		\right) = \frac{t^2\sigma_2^2}{30} \operatorname{diag}\!\left(k_{2z}^2,k_{2z}^2,3k_{2z}^2,\frac{8k_{3x}^2}{3},k_{3x}^2,0\right) \eqqcolon \mymat{D}\ .
	\end{equation}
	Note that the diagonal entries do not represent the eigenvalues of $\mymat{H}$, nor is $\mymat{T}$ an orthogonal matrix. This is then applied to $Q_f$ as a coordinate transformation $\myvec{q}\mapsto \myvec{q}\eqqcolon \mymat{T}\myvec{q}'$, which leaves the last coordinate $q_{6}'=q_6=q_{13z}$ unchanged.
	
	The desired form of $Q_f$ requires further transformations. This follows from the observation that, after applying $\mymat{T}$ to $Q_f$, the form can be abstractly written in terms of the partially new variables $\myvec{q}'$ as
	\begin{equation}
		\sum_{a=1}^5 D_{a a}\cdot (q'_a)^2 + g^{(4)}\left(q'_{6}\right) + \sum_{a=1}^5 q'_{a}\cdot h_{a}^{(3)}(\myvec{q}')\ ,
	\end{equation}
	where the $D_{aa}$ are components of the matrix $\mymat{D}$ from before, $g^{(4)}\in\C\{x\}$ with $\operatorname{ord}(g)\geq 4$ and $h^{(3)}_a\in\C\{x_1,\dots,x_6\}$ with $\operatorname{ord}(h^{(3)}_a)\geq 3$ for $a\in\{1,\dots,5\}$. Only the third term is problematic as it does not separate $q'_6$ from the rest of the variables. Applying the transformation (slightly modified from the original reference to account for $\mymat{T}$ with $\det(\mymat{T})=1$)
	\begin{equation}
		\begin{split}q'_a\ \, &\mapsto q'_a \eqqcolon q''_a - \frac{1}{2 D_{aa}}h^{(3)}_a(\myvec{q}'')   \quad\text{for } a\in\{1,\dots,5\}\ ,\\
			q'_{13z} &\mapsto q''_{13z} \coloneqq q'_{13z}\ . \end{split} \label{eq:splittinglemma:subsequent_trafo}
	\end{equation}
	pushes the mixed term into higher degrees,
	\begin{equation}
		\sum_{a=1}^5 D_{aa}\, (q''_a)^2 + g^{(4)}(q''_{6}) + g^{(6)}(q_{6}) + \sum_{a=1}^5 q''_{a}\cdot h_{a}^{(5)}(\myvec{q}'')\ ,
	\end{equation}
	with $g^{(6)}\in\C\{x\}$ with $\operatorname{ord}(g)\geq 6$ and $h^{(5)}_a\in\C\{x_1,\dots,x_6\}$ with $\operatorname{ord}(h^{(5)}_a)\geq 5$ for $a\in\{1,\dots,5\}$. Reapplying another transformation based on this form continues this process, which terminates due to the prerequisite of the critical point being isolated. Note that all subsequent transformations are the identity to leading order. 
	
	In a final step, note that all convergent power series with a non-zero constant term are invertible. As the residual term $g^{(4)}$ has a non-zero term of degree four, it is possible to write 
	\begin{equation}
		g(q_{6}) \coloneqq \sum_{n=2}^\infty g^{(2n)}(q_6) = c \cdot q_{6}^4 \cdot u(q_{6})
	\end{equation}
	with $c\in\R$  and $u\in\C\{x\}$ with $g(0)=1$. Then the coordinate change
	\begin{equation}
		\begin{split}q''_a\ \, &\mapsto q''_a  \quad\text{for } a\in\{1,\dots,5\}\ ,\\
			q''_{6}=q_6 &\mapsto \tilde{q}_{6} \eqqcolon  \sqrt[\leftroot{2}4]{u(q_6)}\cdot q_6 \end{split} \label{eq:splittinglemma:final_trafo}
	\end{equation}
	removes $u$ completely. The Jacobian determinant of this transformation is unity to first order because the constant term of $u$ is unity. Thus all the transformations leave the integral measure invariant to leading order. In the end, this leaves five variables in $Q_f$ which appear quadratic, and only one with a quartic dependence.
	
	As all the transformations are explicitly given, they can be performed on $Q_f$ in appropriately truncated form. This is done here with the help of Mathematica \cite{mathematica}, especially to verify the cancellations at higher order.
	
	The final result for $Q_f$ after applying the modified splitting lemma is
	\begin{equation}
		\begin{split}Q_f &\approx \frac{t^2\sigma_2^2}{30}\left(k_{2z}^2\left(\myvec{q}_{12}^2+2q_{12z}^2\right) + \frac{k_{3x}^2}{3} \left(8q_{13x}^2 + 3q_{13y}^2\right)\right) \\ &+ \frac{t^2\sigma_3^2}{280}\frac{q_{13z}^4}{k_{2z}^2} \left(k_{3x}^4+6k_{3x}^2k_{3z}^2+
			k_{2z}^2(k_{3x}^2+5k_{3z}^2)+ 5k_{3z}^4 + 2k_{2z}(3k_{3x}^2 k_{3z} + 5k_{3z}^3) \right) 
		\end{split}
	\end{equation}
	as $\myvec{q} \rightarrow 0$. All variables $\myvec{q}''$ have been implicitly renamed to their former names. Taking effects into account only to leading order, the phase transforms only under $\mymat{T}$ as
	\begin{equation}
		k_{2z}q_{12z}+k_{3x}q_{13x}+k_{3z}q_{13z} \ \mapsto\ k_{2z}q_{12z} + \frac{2}{3}\, k_{3x}q_{13x}\ .
	\end{equation}
	
	In this form, the transformed variables can be integrated over with standard techniques because all integration variables are decoupled. The final result is
	\begin{equation}
		\bispec_{k\rightarrow\infty}(\myvec{k}_{2},\myvec{k}_{3}) = \frac{c_0\cdot\expr{-\frac{15}{4}\left(t\,\sigma_2\right)^{-2}}}{\left(t\abs{\sigma_2}\right)^5 \sqrt{t\abs{\sigma_3}}\sqrt[\leftroot{2}4]{c_4(\myvec{k}_2,\myvec{k}_3)}\:k_{3x}^2\abs{k_{2z}}^{\frac{5}{2}}}\ , \label{eq:bispec:splittingresult}
	\end{equation}
	valid for large $k$, with the numerical coefficient $c_0$ and the quartic polynomial $c_4(\myvec{k}_2,\myvec{k}_3)$ given by
	\begin{align}
		c_0 &= 900\sqrt{3}\,\sqrt[\leftroot{2}4]{7000}\: \pi^{\frac{5}{2}}\, \Gamma\!\left(\frac{5}{4}\right) \approx 2.2609 \cdot 10^5\ , \\
		c_4(\mathbf{k}) &= k_{3x}^4+6k_{3x}^2k_{3z}^2+5k_{3z}^4+k_{2z}^2 \left(k_{3x}^2+5k_{3z}^2\right) +2k_{2z}\left(3k_{3x}^2k_{3z}+5k_{3z}^3\right)\ .
	\end{align}
	This formula gives a concise description for the free bispectrum at large $k$, but it only represents the leading order. Higher orders will be dealt with in the next Section. 
	
	The power-law scaling is more apparent in the equilateral configuration $\myvec{k}_2 = k\cdot \hat{\myvec{e}}_z$, $\myvec{k}_3=\frac{\sqrt{3}k}{2}  \hat{\myvec{e}}_x - \frac{k}{2}  \hat{\myvec{e}}_z$
	\begin{equation}
		\bispec^\text{equi}_{k\rightarrow\infty}(k) = \frac{4\sqrt[\leftroot{2}4]{2}\,c_0\,\expr{-\frac{15}{4}\left(t\,\sigma_2\right)^{-2}}}{3\left(t\abs{\sigma_2}\right)^5 \sqrt{t\abs{\sigma_3}}\,k^{\frac{11}{2}}}\ . \label{eq:bispec:splittingresult:equi}
	\end{equation}
	This scaling behaviour extends to other triangle configurations with fixed angles and varying size $k$.

	\subsection{Asymptotics Beyond Leading Order} \label{sec:methods:higherorder}
	
	Having an asymptotic expansion beyond leading order is helpful for several reasons. Firstly, this will improve the accuracy of the asymptotic expansion, seen as an approximation to the full expression, in the transition regime towards large $k$. Secondly, this will give an estimate of the expected error of the approximation, and with this, an expected domain of validity.
	
	In the last Section, higher orders might have been taken into account on several occasions:
	\begin{enumerate}
		\item The first transformation based on the Hessian is exact and no approximations were used.
		\item The subsequent transformations are the identity only to leading order. Their Jacobian effectively adds a polynomial in front of the exponentials.
		\item The subsequent transformations transform the complex phase as well.
		\item The final transformation based on $g(q_{13z})$ can be taken into account beyond leading order.
	\end{enumerate}
	In principle, the second and third points could make it impossible to calculate any higher orders, as arbitrarily many transformations would have to be taken into account. However, the transformations in Eq.~(\ref{eq:splittinglemma:subsequent_trafo}) will always have a part linear in $q$, but the next order will increase with each successive transformation.
	
	We will now calculate the next two orders in $k$. The details can be found in Appendix~\ref{sec:app:bispec}. The resulting expression for general $\myvec{k}$ configurations is long, thus we leave this to the Appendix. For an equilateral $\myvec{k}$ configuration, the next-order result is
	\begin{equation}
		\bispec_{k\rightarrow\infty}^\text{NO,equi}(k) \approx \frac{32 \sqrt[\leftroot{2}4]{8\cdot 7^2}\,c_0^\text{NO}}{3} \frac{ \expr{-\frac{15}{4}\left(t\,\sigma_2\right)^{-2}} f_\text{equi}^\text{NO}(k,\sigma,t)}{t^9\,\abs{\sigma_2}^{11}\, k^7 \sqrt{t\abs{\sigma_3}\, k}}\ ,
	\end{equation}
	with the function $f_\text{equi}^\text{NO}$ encoding the different $k$ orders 
	\begin{equation}
		f_\text{equi} = \sqrt{35}\,\Gamma\!\left(\frac{9}{4}\right)\!\left(2688\,t^4k^2\sigma_2^6 + 5\sigma_3^2(3582\,\sigma_2^2 t^2-5995)\right) - 2400\: \Gamma\!\left(\frac{11}{4}\right) t^3\, k\, \sigma_2^2\,\abs{\sigma_3}
	\end{equation}
	and the constant
	\begin{equation}
		c_0^\text{NO} = \frac{15\sqrt{3}\,\sqrt[\leftroot{2}4]{5}\: \pi^{\frac{5}{2}}}{224\,\sqrt[\leftroot{2}4]{14^3}}\approx 0.419204\ . 
	\end{equation}
	This reveals that three orders in $k$ occur, namely $k^{-15/2}$, $k^{-13/2}$, and $k^{-11/2}$. The result thus incorporates two corrections to the leading-order expression. The different orders of the free bispectrum are called $\bispec^{(0)}_{k\rightarrow\infty}$, $\bispec^{(1)}_{k\rightarrow\infty}$, and $\bispec^{(2)}_{k\rightarrow\infty}$, together summing up to $\bispec^\text{NO}_{k\rightarrow\infty}$.
	
	The $k$ where the zeroth and first order meet in the equilateral configuration is given by
	\begin{equation}
		k^{0=1}_\text{equi} = \frac{5\sqrt{5}(56 \sigma_2^4\sigma_4^2 t^2 + 27 \sigma_3^4 (2\sigma_2^2 t^2 - 15))\Gamma\!\left(\frac{7}{4}\right)}{864\sqrt{7} \sigma_2^4 \sigma_3^{3}t^3\Gamma\!\left(\frac{9}{4}\right)}\approx 7.06\, h\operatorname{Mpc}^{-1}\ .
	\end{equation}
	Intersections like this can be calculated for any given triangle configuration and give an expected wave number from which onwards the asymptotic description should be applicable.
	
	\subsection{Asymptotics for Degenerate Configurations} \label{sec:methods:parallel}
	
	We will finally analyze degenerate configurations of the $\myvec{k}$ vectors of the free bispectrum, i.e., triangle configurations with zero surface area. The cases of either one of the wave vectors $\myvec{k}_i$ being entirely zero will not be treated here, as this requires an analysis of the neglected shot-noise contributions from Eq.~(\ref{eq:bispec:generalderivation}).
	
	Then only one degenerate case remains: Taking $k_{3x}=0$, the vectors $\myvec{k}_2$ and $\myvec{k}_3$, and thus $\myvec{k}_1$, become parallel. The only remaining relevant components are $k_{2z}$ and $k_{3z}$. The restriction $k_{2z}+k_{3z}\neq 0$ must be imposed, however, otherwise $\myvec{k}_1$ would vanish.
	
	This case is important to examine because the asymptotic description in previous Sections breaks down for parallel configurations, indicated by the divergence of Eq.~(\ref{eq:bispec:splittingresult}) when setting $k_{3x}=0$. As argued in the accompanying discussion, originally in \cite{bern02}, an increase in amplitude of the bispectrum for parallel configurations means that such configurations contribute more in terms of structure formation. This shape dependence will be investigated further in the following Section.
	
	In this $\myvec{k}$ configuration, the Hessian matrix $\mymat{H}$ at the critical point zero is the same as before, but with $k_{3x}=0$. Its rank and corank are then $3$, which is somewhat problematic, as the left-over polynomial $g$ in the splitting lemma becomes significantly more complicated. The critical points are never simple for corank 3 and thus easy normal forms exist only for special cases \cite{greu07, arno74}. Nevertheless, the transformations in the proof of the splitting lemma turn out to be useful in this approximate, asymptotic setting.
	
	In principle, the same steps have to be followed as before. The details can be found in Appendix~\ref{sec:app:bispecdegen}. The result has the form
	\begin{equation}
		\bispec^{\parallel}_{k\rightarrow\infty}(\myvec{k}_{2},\myvec{k}_{3}) \approx \frac{c_0^\parallel \cdot\expr{-\frac{5}{2}\left(t\,\sigma_2\right)^{-2}}}{\left(t\abs{\sigma_2}\right)^3 \abs{t\,\sigma_3\, k_{2z}\,k_{3z}(k_{2z}+k_{3z})}^{\frac{3}{2}}} \ , \label{eq:bispec:parallel}
	\end{equation}
	with the numerical coefficient
	\begin{equation}
		c_0^\parallel = -150\sqrt[\leftroot{2}4]{5\cdot 14^3}\: \pi^{\frac{5}{2}}\, \Gamma\!\left(-\frac{1}{4}\right) \operatorname{_2F_1}\!\left(\frac{1}{2},\frac{3}{4} ;\frac{3}{2},-4 \right)\approx 87\,400\ .
	\end{equation}
	To see the scaling behaviour and to make the connection to non-parallel configurations, define a special one: take $\myvec{k}_2$ and $\myvec{k}_3$ to be of the same length and pointing into the same direction. Call this new length again $k$, then simply $k_{2z}=k$, $k_{3z}=k$, and $k_{1z}=-2k$ holds. In this case, the free bispectrum becomes
	\begin{equation}
		\bispec^{\parallel\text{,special}}_{k\rightarrow\infty}(k) \approx \frac{c_0^\parallel}{2\sqrt{2}} \frac{\expr{-\frac{5}{2}\left(t\,\sigma_2\right)^{-2}}}{\left(t\abs{\sigma_2}\right)^3 \abs{t\,\sigma_3}^{\frac{3}{2}}\, \abs{k}^{\frac{9}{2}}}\ .
	\end{equation}
	Compared to Eq.~(\ref{eq:bispec:splittingresult:equi}), this shows a power-law scaling with $k$ of only $-9/2$ rather than $-11/2$ before, which is a significant increase in overall amplitude. In similar ways, the scaling with respect to $t$ is $t^{-9/2}\exp(-t^{-2})$. Also the powers of the $\sigma$ have been rearranged, where now it is $\sigma_2^{-3}\exp(-\sigma_2^{-2})\sigma_3^{-3/2}$, to before $\sigma_2^{-5}\exp(-\sigma_2^{-2})\sigma_3^{-1/2}$.
	
	\subsection{Comment on Higher N-Point Correlation Functions}
	
	The asymptotic calculations performed here, based on the splitting lemma, can also be applied to higher $n$-point correlation functions, such as the trispectrum. The corank of the Hessian is then larger than $1$, making computations more complicated. Also, there are more possibilities to form degenerate $\myvec{k}$ vector configurations which then need separate investigations. In the case of the trispectrum, we were still able to calculate the leading order asymptotic terms.
	
	\subsection{Numerical validation}
	
	Having used the splitting lemma to derive the asymptotic limit of the free bispectrum for large $k$, we proceed to numerically evaluate the full expression for the free bispectrum with a focus on small scales. For this purpose, we wrote code in \verb!C++!$\,$\footnote{This code can be found in this repository: \href{https://lin0.thphys.uni-heidelberg.de:4443/kft-bispec/bispec-asymp-paper}{GitLab ITP}.}, using the GNU scientific library (GSL) \cite{gsl}. 
	
	The final form of the free bispectrum is given in Eq.~(\ref{eq:bispec:general}), which can schematically be written as
	\begin{equation}
		\bispec(\myvec{k}_2, \myvec{k}_3) = \int_{q_{12}} \int_{q_{13}}  \expr{-Q_{D}}\, \expr{-t^2 Q_{C}}\, \expr{\iu\cdot\phi}
	\end{equation}
	with the damping $Q_{D}$, the quadratic form free of damping $Q_{C}$, and the Fourier phase $\phi$ defined in Eq.~(\ref{eq:bispec:quadform1}) and Eq.~(\ref{eq:bispec:quadform2}). The triangle condition for the $\myvec{k}$ is implicit here.
	
\begin{figure*}
		\includegraphics[width=\textwidth]{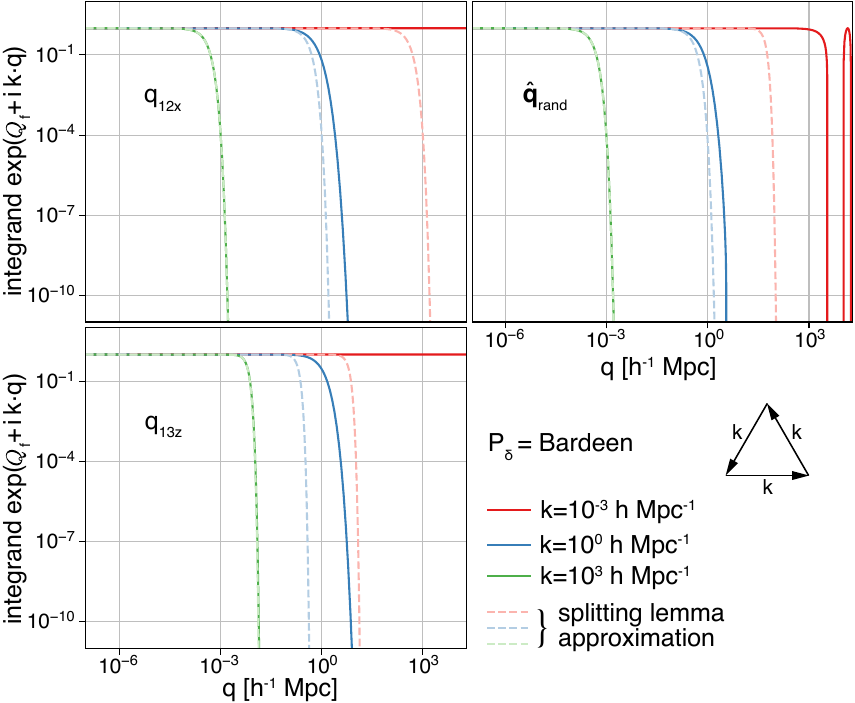}
		\caption{The absolute value of the integrand of the bispectrum (cf.\ Eq.~(\ref{eq:bispec:general})) is shown in the three panels. In the top left one, the integrand is plotted along the $q_{12x}$ axis which is non-oscillating. The bottom panel shows the integrand along the $q_{13z}$ axis, on which the behaviour is expected to go with $\sim\exp(-x^4)$. The right panel shows the integrand along a random direction in $\myvec{q}$-space.\\
			The Bardeen power spectrum \cite{bard86} and an equilateral $\myvec{k}$ configuration are used. The integrand has been transformed based on a transformation of the Hessian of the exponent of the integrand of the bispectrum. The dashed lines show the integrand's asymptotic approximation by the splitting lemma. For large triangle sidelengths $k$, the agreement of the approximation and the original integrand can be seen in all directions.}
		\label{fig:integrand:trafo}
	\end{figure*}

	This is in principle a six-dimensional integration over an oscillating integrand, which is numerically challenging. Initial tests showed that in this form, the slight differences between a quadratic and quartic Gaussian could not be seen well enough. Thus, the linear transformation of coordinates into a diagonal system for the Hessian of the quadratic form was applied. As the change of coordinates is linear, it can be implemented exactly. The transformed integrand can be seen in Fig.~\ref{fig:integrand:trafo}. The figure also shows the prediction derived from the splitting lemma as a dashed line.
	
	Calculating $Q_C$ needs the momentum auto-correlations. These can be calculated completely from the functions $a_1(q)$ and $a_2(q)$. Asymptotic descriptions exist, but the function values need to be available in all regimes. The code interpolates the function values across the necessary domain, so that the workload is reduced drastically in the subsequent numerical integration. The code also uses the asymptotic description in the appropriate limits, as the numerical integration becomes unstable for small values of $q$. In the asymptotic regime, the series was evaluated up to fourth order, i.e., up to the $q^8$ term. The result can be seen in Fig.~\ref{fig:integrand:a1a2}.
	
	Due to the high dimension of the integration, the Monte-Carlo VEGAS algorithm in its GSL implementation was chosen as it produces consistently better results than the MISER and plain variants. To ensure absolute convergence of the integrand, several Taylor orders were subtracted from the exponential as in
	\begin{equation}
		\expr{-Q_{D}} \left( \expr{-t^2 Q_{C}} - 1 + t^2 Q_{C} - \frac{1}{2} \left(t^2 Q_{C}
		\right)^2\right) \cos(\phi)\ .
	\end{equation}
	We have already used here that the integrand represents an overall even function, so the imaginary part of the phase cancels. This procedure has almost no effect for large $k$ values, but it severely damps the integrand for low $k$ as this procedure effectively just subtracts the small $k$ approximation of the bispectrum shown before in Eq.~(\ref{eq:bispec:largescale}). As the analytical result of the integration of these Taylor orders is already well-known, it can be restored after numerical integration.
	
	While the integration domain is unbounded, numerical integration cannot deal with it as is. The integrand needs to fall off sufficiently quickly, otherwise the integral would be divergent. Two options exist then: one can either map the infinite domain into a finite one by coordinate transformation, or integrate only up to a finite boundary that is set such that the integrand is so small that the remaining contribution to the integral is negligible. Here the latter option was chosen. Setting these limits correctly is important. If chosen too large, the density of sampled points would drop quickly, requiring an extraordinary amount of computing power. Additionally, the central peak at zero is important, but the volume of its support can quickly vanish in relation to the total integration volume, as the volume increases rapidly in six-dimensional space. However, setting the volume too small cuts off important parts of the function.
	
	With this in mind, the following system for setting the integration domain was implemented. The idea was that, in the large $k$ limit, the integrand can be well described by a Gaussian.  When this does not apply, the oscillations can be used to define a point after which the integral probably cancels.
	
	We thus introduce two parameters, $p_\text{fall-off}$ and $p_\text{max. osc.}$. The first is set to a level relative to the central peak of the integrand at zero, after which the integrand can be negligibly small. We mostly set $p_\text{fall-off}=10^{-4}$. The second parameter describes after how many oscillations the integration can be assumed to cancel itself. We mostly used $p_\text{max. osc.}=12$, which produced stable results.
	
	Note that the application of the linear transformation put the quartic Gaussian on the $q_{13z}$ axis, and that it is now possible to set specific limits for it, which would have been significantly harder without the transformation. Using all this sets the upper bound for integration. The lower bound is simply chosen as the negative of the upper bound.
	
	Generally, the code writes all the choices and parameter values into a log-file, so close monitoring of the process is possible and adjustments can then be made on an informed basis. More detail on the parameters can be found in Appendix~\ref{sec:app:numerics}. Many of the choices in this section are due to the interest in the large $k$ behaviour. Other regimes, e.g. around the onset of non-linearities need a different numerical treatment to produce precise results \cite{Chen_2024}.
	
	\section{Results} \label{sec:results}
	
\begin{figure*}
		\includegraphics[width=\textwidth]{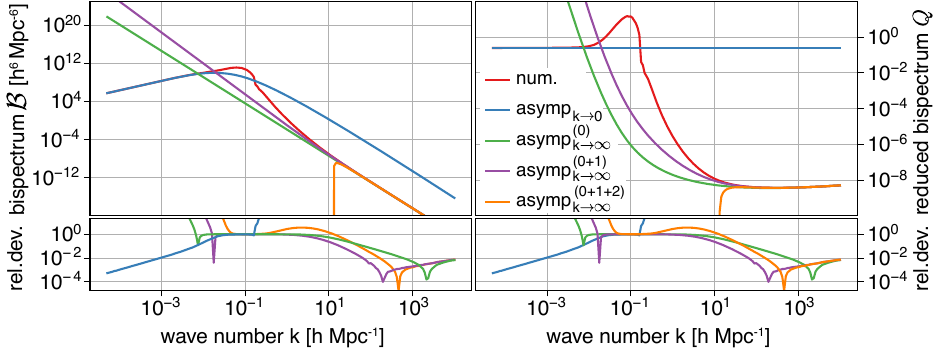}
		\caption{The left panel shows the numerical integration result of the free bispectrum, with asymptotic approximations for $k\rightarrow 0$ and $k\rightarrow\infty$. The right panel shows the reduced bispectrum, where the bispectrum is normalized with the square of the linearly evolved power spectrum. The bottom figures show the relative deviation with respect to the numerical result. The $\myvec{k}$ vector configuration used is an equilateral triangle with side length $k$. The initial power spectrum is taken as the Bardeen power spectrum \cite{bard86} using an exponential cut-off with $k_s=10^2 \, h\operatorname{Mpc}^{-1}$.}
		\label{fig:bispec:result}
	\end{figure*}

	The main result of this paper is the asymptotic behaviour of the free bispectrum with initial momentum-momentum correlations, i.e., of the Zel'dovich approximation bispectrum. Its analytic calculation was confirmed by direct numerical computation. The result can be seen in Fig.~\ref{fig:bispec:result}. There the typical behaviour of asymptotic series become apparent: while providing a better approximation in a certain regime, the farther the point is from the expansion point, the worse the higher-order approximation gets. This is because in general, asymptotic series tend to be divergent. 
	In the small-scale regime, the relative deviation between numerics and small-scale asymptotics seems to level off, which is most probably due to issues in the numerical integration. In the small-scale regime of the reduced bispectrum $Q$ it is important to see that the curve does not level off at a constant value as expected from large-scale approximations \cite{bern02, scoccimarro2001fitting, fry1984galaxy}, but retains a residual $k^{1/2}$ dependence due to the non-perturbative degeneracy. 
	
\begin{figure*}
		\includegraphics[width=\textwidth]{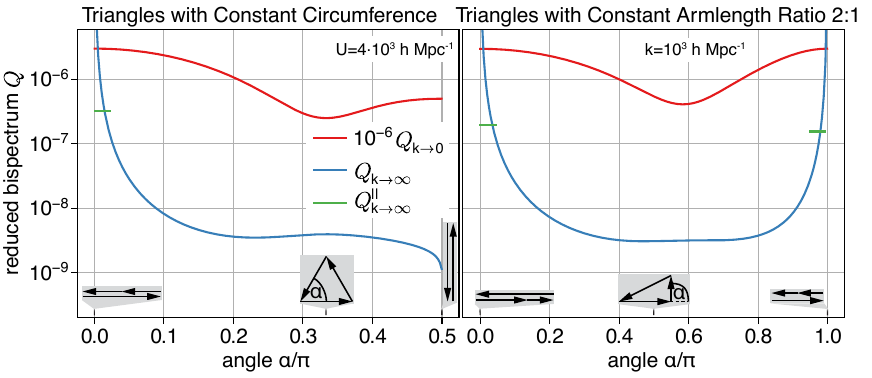}
		\caption{Both panels show the dependence of the reduced bispectrum $Q$ on the $\myvec{k}$ configuration. The left panel is an isosceles configuration, with changing angle $\pi-\alpha$ between $\myvec{k}_2$ and $\myvec{k}_3$ while keeping the total circumference $U=\abs{\myvec{k}_1}+\abs{\myvec{k}_2}+\abs{\myvec{k}_3}=4000 h\text{Mpc}^{-1}$ constant. The right panel keeps $\abs{\myvec{k}_2}$ and $\abs{\myvec{k}_3}$ constant, with $\abs{\myvec{k}_2}=2 \abs{\myvec{k}_3} = 1000h\text{Mpc}^{-1}$. The reduced bispectrum in the large-scale approximation $Q_{k\rightarrow 0}$ is rather insensitive to cosmology, specifically there is no dependence on time or overall $k$-scale in this lowest order approximation \cite{bern02}. The values of the large-scale approximation had to be rescaled such that both results fit in the same plot.}
		\label{fig:bispec:angulardep}
	\end{figure*}

	Another interesting feature to investigate is the dependence of the bispectrum on the shape of the $\myvec{k}$-vector configuration. This can be found in Fig.~\ref{fig:bispec:angulardep}.
	Here, two special parametrizations are used:
	\begin{enumerate}
		\item An isosceles configuration with constant circumference. To enforce the isosceles condition, set $k_{3z}=-\frac{k_{2z}}{2}$. This then implies, using the angle $\pi-\alpha$, $\alpha\in[0,\frac{\pi}{2}]$ between $\myvec{k}_2$ and $\myvec{k}_3$ and the total circumference $U=\abs{\myvec{k}_1}+\abs{\myvec{k}_2}+\abs{\myvec{k}_3}$
		\begin{equation}
			k_{2z} = \frac{\cos(\alpha)}{1+\cos(\alpha)}\, U\ , \ \ k_{3x} = \tan\!\left(\frac{\alpha}{2}\right) \frac{U}{2} \ .
		\end{equation}
		Note that the limit $\alpha=0$ can be described by the parallel configuration result while $\alpha=\frac{\pi}{2}$ cannot, as there $\myvec{k}_2=0$. Schematically, this can be draw as ($\myvec{k}_1$ is completely determined through the $\diracd{-}$ triangle condition):
		\begin{center}
			\begin{tikzpicture}[>=stealth]
				\draw[->] (0,0) -- (2,0);
				\draw[->] (2,0) -- (1,1.73);
				\draw[->] (1,1.73)-- (0,0);
				\draw (1,-0.3) node {$\myvec{k}_2$};
				\draw (1.75,1.15) node {$\myvec{k}_3$};
				\draw (0.2,1) node {$\myvec{k}_1$};
				\draw (1.65,0.2) node {$\alpha$};
				\draw ([shift=(120:0.6)]2.0,0.0) arc (120:180:0.6);
				\draw (2.25,0.42) node {$\pi\!-\!\alpha$};
				\draw ([shift=(0:0.85)]2.0,0.0) arc (0:120:0.85);
				\draw[-,dashed] (2.0,0.0)-- (2.85,0);
			\end{tikzpicture}
		\end{center}
		\item A configuration in which $\abs{\myvec{k}_2}$ and $\abs{\myvec{k}_3}$ are kept constant. Furthermore, the condition $\abs{\myvec{k}_2}=2\abs{\myvec{k}_3}$ is imposed. The configuration can then be parameterized by $k=\abs{\myvec{k}_2}$ and the angle $\alpha\in[0,\pi]$ is between $\myvec{k}_2$ and $\myvec{k}_3$. This can then be described by
		\begin{equation}
			k_{2z}=k\ , \ \ k_{3x} = \sin(\alpha)\frac{k}{2}\ , \ \ k_{3z}=\cos(\alpha)\frac{k}{2}\ .
		\end{equation}
		Note that here both extremal cases $\alpha=0$ and $\alpha=\pi$ can be described via the formula (\ref{eq:bispec:parallel}) for $\bispec_{k\rightarrow\infty}^\parallel$. Schematically, this can be draw as ($\myvec{k}_1$ is completely determined through the triangle condition):
		\begin{center}
			\begin{tikzpicture}[>=stealth]
				\draw[->] (0,0) -- (2,0);
				\draw[->] (2,0) -- (2,1);
				\draw[->] (2,1)-- (0,0);
				\draw (1,-0.3) node {$\myvec{k}_2$};
				\draw (1.75,0.4) node {$\myvec{k}_3$};
				\draw (0.95,0.8) node {$\myvec{k}_1$};
				\draw (2.23,0.23) node {$\alpha$};
				\draw ([shift=(0:0.6)]2.0,0.0) arc (0:90:0.6);
				\draw[-,dashed] (2.0,0.0)-- (2.6,0);
			\end{tikzpicture}
		\end{center}
	\end{enumerate}
	Especially the second configuration has been used extensively in the literature \cite{bern02,scoccimarro2001fitting,hivon1994redshift}. 
	
	Looking at the case of varying angle, but keeping the circumference constant, the isosceles configuration forms a local maximum. Analytical calculations show as well that this occurs at $\alpha=\frac{\pi}{3}$. In the bispectrum this is actually a global minimum, but turns into a local maximum in the reduced bispectrum. With the interpretation from \cite{lewi11}, this means that in the \Zel\ picture at this wave number, filaments with a circular cross section are slightly more abundant than those with elliptical ones. Generally, as the overall sign of the free bispectrum is positive, this indicates that the cosmos is filled with higher overdensities and rather mild, extended underdensities \cite{lewi11}. In the constant armlength configuration, a minimum can also be calculated, but a specific interpretation for the angle is not known to the authors. 
	
	In both configurations, the reduced bispectrum for parallel configurations is an order of magnitude higher than the typical value of $Q$ for in the middle of the $\alpha$-range. Additionally, the large-scale approximation leads to a reduced bispectrum that is far more insensitive to cosmology and time \cite{bern02,fry1984galaxy} than the small-scale result, which explitly depends on the overall scale and time, and on cosmological parameters through the moments of the initial density power spectrum.
	
\begin{figure*}
		\includegraphics[width=\textwidth]{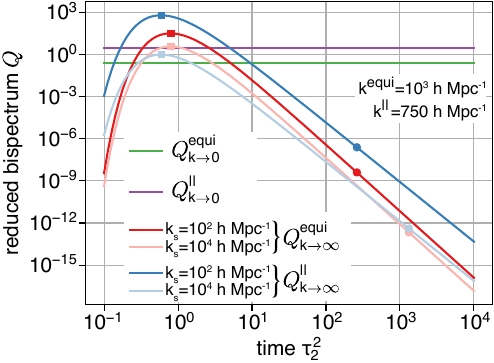}
		\caption{The figure gives the value of the reduced bispectrum for two configurations as a function of time $\tau_2^2=t^2\sigma_2^2$. The first configuration is the equilateral one, corresponding to cosmic filaments. The second one is a parallel one, corresponding to pancakes \cite{lewi11}. The large-scale approximation, to lowest order, is completely insensitive to the used cosmological model and time, in these $\myvec{k}$ configurations. The small-scale asymptotics gives an interesting growth of different structures within the \Zel\ picture. The dots in the plot indicate what would correspond to the current cosmic time, given different values of $\sigma_2^2$ due to different cut-offs $k_s$. The squares indicate the maximum amplitude, which are shown to be independent of the higher moments $\sigma_{n>2}^2$.}
		\label{fig:bispec:timedep}
	\end{figure*}

	Finally, following the analysis of time dependence in \cite{konrad2022kinetic}, define the time 
	\begin{equation}
		\tau_2^2=t^2 \cdot \sigma_2^2\ .
	\end{equation}
	The time dependence of the power spectrum could be fully expressed by this coordinate in the small-scale regime. For the free bispectrum, due to the degeneracy in the critical point, even to lowest order $\sigma_3^2$ plays a relevant role. Figure~\ref{fig:bispec:timedep} shows the time dependence of the small- and large-scale approximations to the reduced bispectrum. 
	
	Two specific configurations are singled out here: the equilateral and the special-parallel ($k=k_{2z}=k_{3z}$) case, both with the same circumference of $3000 h\text{Mpc}^{-1}$. Using the interpretation of \cite{lewi11}, these correspond to filaments and pancakes. Note that for these configurations, the large-scale approximations become completely insensitive to the power spectrum used, as that divides out. Thus also all time dependence is lost in this lowest-order approximation.
	
	The small-scale approximation shows interesting behaviour however. Just as expected from the eigenvalue distribution of the \Zel\ tensor \cite{sheth2001ellipsoidal}, the collapse of the homogeneous background happens first along one axis, leading to pancake-like structures, which can then further collapse to filaments \cite{zel1970gravitational}. This is reflected in the overall higher values for the parallel configuration and its steeper rise. 
	
	Analytical calculations also show that the position of the maxima are independent of the higher moments, and given by
	\begin{equation}
		\operatorname{argmax}_{\tau_2^2}\left(Q_{k\rightarrow\infty}^\text{equi}(\tau_2^2)\right) = \frac{15}{11}\ , \quad 
		\operatorname{argmax}_{\tau_2^2}\left(Q_{k\rightarrow\infty}^\parallel(\tau_2^2)\right) = \frac{10}{9}\ .
	\end{equation}
	
	These are universal with respect to the initial power spectrum and cut-off used, in the sense that the only dependence is in $\sigma_2^2$, which is absorbed into the time coordinate $\tau_2^2$. Again, the parallel configurations reaches its maximum earlier, as expected from the \Zel\ model.
	
	\section{Summary and Outlook} \label{sec:summary}
	
	The main result of this paper is the asymptotic behaviour of the free bispectrum, i.e., the bispectrum within the \Zel\ approximation. This is based on the following assumptions:
	\begin{itemize}
		\item The initial density perturbation field is a Gaussian random field with an appropriate small-scale behaviour, i.e., its moments converge;
		\item The cosmic time is late enough for the momentum-momentum correlations to dominate density-density and density-momentum correlations;
		\item The wave numbers are sufficiently large. This can be estimated through the intersection of the leading order and next-to-leading order asymptotics.
	\end{itemize}
	Then the following formula shows the universal form of the free bispectrum for large $k$
	\begin{equation}
		\bispec_{k\rightarrow\infty}(\myvec{k}_{2},\myvec{k}_{3}) = \frac{c_0\cdot\expr{-\frac{15}{4}\left(t\,\sigma_2\right)^{-2}}}{\left(t\abs{\sigma_2}\right)^5 \sqrt{t\abs{\sigma_3}}\sqrt[\leftroot{2}4]{c_4(\myvec{k}_2,\myvec{k}_3)}\:k_{3x}^2\abs{k_{2z}}^{\frac{5}{2}}}\ ,
	\end{equation}
	with the numerical coefficient $c_0$ and the quartic polynomial $c_4(\myvec{k}_2,\myvec{k}_3)$ given by
	\begin{align}
		c_0 &= 900\sqrt{3}\,\sqrt[\leftroot{2}4]{7000}\: \pi^{\frac{5}{2}}\, \Gamma\!\left(\frac{5}{4}\right) \approx 2.2609 \cdot 10^5\ , \\
		c_4(\mathbf{k}) &= k_{3x}^4+6k_{3x}^2k_{3z}^2+5k_{3z}^4+k_{2z}^2 \left(k_{3x}^2+5k_{3z}^2\right) +2k_{2z}\left(3k_{3x}^2k_{3z}+5k_{3z}^3\right)\ .
	\end{align}
	Higher-order corrections are derived as well in Sect.~\ref{sec:methods:higherorder}. This formula holds for triangle configurations of the $\myvec{k}$ vectors that are non-degenerate, i.e., have $k_{3x}\neq 0$ and all $\myvec{k}$ non-zero. A similar result for parallel configurations was derived as well in Sect.~\ref{sec:methods:parallel}.
	
	This asymptotic result was then confirmed by direct numerical computation of the free bispectrum\footnote{This implementation and other computations are available at: \href{https://lin0.thphys.uni-heidelberg.de:4443/kft-bispec/bispec-asymp-paper}{GitLab ITP}.}. Looking at specific configurations with only one scaling parameter $k$ gives the following behaviour:
	\begin{itemize}
		\item For all components $k_{3z}$, $k_{3x}$, $k_{3z}$ large, the overall power law exponent is $-11/2$;
		\item For parallel configurations with $\myvec{k}_1\neq 0$ and $k_{2z}$, $k_{3z}$ large, the overall power law exponent is $-9/2$.
	\end{itemize}
	Using the interpretation of \cite{lewi11} of what these structures look like in real space allows to identify the connection of equilateral triangles with filamentary structures, and specific parallel configurations with pancakes. We then investigated the growth of these different types of structure and found an early preference for pancake-like structures, as expected from the \Zel\ model of gravitational collapse. Additionally, there was a richer dependence on the initial power spectrum, compared to the \Zel\ power spectrum in the small-scale limit: its dependence on $\sigma_3^2$ to leading order due to the non-perturbative degeneracy means that there is also a richer dependence on cosmological parameters and the initial cut-off scale $k_s$ used for the initial power spectrum (\ref{eq:spectral:cutoff}). 
	
	We also investigated the overall shape dependence of the reduced bispectrum in this small-scale approximation. We found that there is a higher residual dependence on cosmology and time compared to lowest-order large-scale computation. We also note here that parallel configurations had a significantly higher amplitude, and that the equilateral configuration forms a small local maximum when varying the angle of the configuration while keeping the circumference constant. 
	
	The result can be extended in two directions: Firstly, other initial correlations beyond momentum-momentum correlations were neglected. Inclusion of those would be in principle interesting, but including them into the \Zel\ power spectrum has shown in other contexts that they are only relevant close to the initial time. Nevertheless, especially for other versions of KFT, such as resummed KFT \cite{lilow2019resummed}, getting this correctly for small $t$ would be important.
	
	Secondly, including more gravitational interactions relative to the \Zel\ trajectories would be very interesting. In this perturbative sense, this result here provides the lowest-order term. Already here we see an interesting effect from treating the initial correlations non-perturbatively, and it would be important to see how this persists when including more interactions. Especially for investigations into virialization this becomes important: When the kinetic energy becomes completely unordered on small scales, the reduced bispectrum is expected to loose all dependence on the shape of the triangle configuration \cite{scoccimarro2001fitting}. Here however, we do not see this behaviour. The reduced bispectrum retains the overall dependence on the wave number $k$ through a residual $k^{1/2}$, as the free bispectrum does not scale with $k^{-6}$ on small scales as expected from large-scale perturbation theory.
	
	\acknowledgments
	
	This work was funded in part by the Deutsche Forschungsgemeinschaft (DFG, German Research Foundation) under Germany's Excellence Strategy EXC 2181/1 - 390900948 (the Heidelberg STRUCTURES Excellence Cluster).
	
	\appendix
	
	\section{Checking the Requirements for the Splitting Lemma} \label{sec:app:checksplit}
	
	This Section shows the calculations on why the splitting lemma can be applied to $Q_f$ in the first place, defined in Eq.~(\ref{eq:bispec:quadformfull}). These are:
	\begin{enumerate}
		\item $Q_f$ has a critical point at $\myvec{q}=0$
		\item $Q_f$ evaluates to zero at $\myvec{q}=0$
		\item The critical point of $Q_f$ in $\myvec{q}=0$ is isolated
		\item $Q_f$ can be written as a convergent power series in a neighbourhood of $\myvec{q}=0$
	\end{enumerate}
	Before point 3, this Section will also provide details on how to calculate the Hessian of $Q_f$ at zero.
	
	\subsection{Critical Point}
	This Section will calculate the derivative of the quadratic form $Q(l)$ for general $l\in\N\setminus\{1\}$. This quadratic form arises when calculating not only the bispectum ($l=3$), but higher $l$-point correlators.
	
	The quadratic form is given by
	\begin{equation}
		Q(l) = \frac{t^2}{2} \frac{\sigma_1^2}{3} \sum_{i=1}^l \myvec{k}_{i}^{2} + t^2 \sum_{i=2}^l \myvec{k}_{1}^\top \mymat{C}_{pp}(\myvec{q}_{1i}) \myvec{k}_{i} + t^2 \sum_{i=2}^{l-1}\sum_{j=i+1}^l \myvec{k}_{i}^\top \mymat{C}_{pp}(\myvec{q}_{1i} - \myvec{q}_{1j}) \myvec{k}_{j}\ . 
	\end{equation}
	The derivative is taken with respect to $q_{1a,m}$, the $m$-th component of $\myvec{q}_{1a}$, for $a\in\{2,\dots,l\}$ and $m\in\{1,2,3\}$. Clearly, the first part of the sum is a constant for the $q$ derivative, so the derivative is given by
	\begin{equation}
		\begin{split}
			\dpart{q_{1a,m}} Q(l) &= t^2 \dpart{q_{1a,m}} \left.{\vphantom{ \sum_{j=a+1}^l}}\right[ \myvec{k}_{1}^\top \:\!\mymat{C}_{pp}(\myvec{q}_{1a})\:\! \myvec{k}_{a}  + \sum_{\substack{i=2\\ i \neq a}}^{l} \myvec{k}_{i}^\top \:\!\mymat{C}_{pp}(\myvec{q}_{1i} - \myvec{q}_{1a})\:\!  \myvec{k}_{a} \left.{\vphantom{ \sum_{j=a+1}^l}}\right] \\
			&= t^2 \dpart{q_{1a,m}}\, \sum_{\substack{i=1\\ i \neq a}}^{l}\, \myvec{k}_{i}^\top \:\!\mymat{C}_{pp}(\myvec{q}_{1i} - \myvec{q}_{1a})\:\!  \myvec{k}_{a} \\
			&=-t^2\sum_{\substack{i=1\\ i \neq a}}^{l} \int_{\myvec{k}} (\myvec{k}_{i}\cdot \myvec{k})(\myvec{k}_{a}\cdot \myvec{k}) P_\psi(k) \sin\!\left(\myvec{k}\cdot(\myvec{q}_{1a}-\myvec{q}_{1i})\right)(\myvec{k})_m\ ,
		\end{split}
	\end{equation}
	where $(\myvec{k})_m$ denotes the $m$-th component of the vector $\myvec{k}$. Going from the second to the third line, it was used that $\myvec{q}_{11}=0$, so the expression could be put neatly into one sum. For the last expression it was used that the Fourier phase can be written as a cosine because $\mymat{C}_{pp}$ is an even function. 
	
	In this form, it is clear to see that for $\myvec{q}_{1a}=0\ \forall\ a\in\{2,\dots,l\}$, the derivative vanishes.
	
	\subsection{Value at Critical Point}
	Based on the form of the correlations $\mymat{C}_{p_i p_j}(\myvec{q})$ in equation~(\ref{eq:cpp:definition}), it is easy to see that
	\begin{equation}
		\mymat{C}_{p_i p_j}(0) = \idmat_3 a_1(0) = \idmat_3 \frac{\sigma_1^2}{3}\ .
	\end{equation}
	Plugging this into the quadratic form gives
	\begin{equation}
		Q(l)|_{\myvec{q}=0} = \frac{t^2}{2} \sum_{i=1}^l \myvec{k}_i \sum_{j=1}^l \mymat{C}_{pp}(0)\, \myvec{k}_j = \frac{t^2 \sigma^2_1}{6} \sum_{i=1}^l \myvec{k}_i \sum_{j=1}^l \myvec{k}_j = 0\ , 
	\end{equation}
	which equates to zero because of homogeneity enforcing the $\diracd{\sum_{i=1}^l \myvec{k}_i}$ condition for the $\myvec{k}$ vectors.
	
	\subsection{Convergent Power series}
	The quadratic form $Q(l)$ can be written as a power series in $\myvec{q}$ with use of the expansions of the $a_1$ and $a_2$ functions given in equations~(\ref{eq:a1:definition})-(\ref{eq:a2:definition}). The series has been shown to be convergent for models with an exponential cut-off in the initial power spectrum $P^\mathrm{(i)}_\delta (k) \expr{-\frac{k}{k_s}}$ \cite{konr20}. This cut-off is needed to make the moments of the initial power spectrum finite.
	
	\subsection{Hessian}
	The Hessian of the quadratic form $Q(l)$ is used multiple times in this thesis. We will again show the result for general $l\in\N\setminus\{1\}$, the bispectrum corresponds to $l=3$.
	
	For $a,b\in\{2,\dots,l\}$, $m,n\in\{1,2,3\}$, the Hessian is given by
	\begin{equation}
		\begin{split}
			\frac{\partial^2\:\!Q(l)}{\partial\:\! q_{1a,m}\:\!\:\!\partial\:\! q_{1b,n}} &=t^2 \dpartsq{q_{1a,m}}{q_{1b,n}} \left[ \sum_{i=2}^l \myvec{k}_{1}^\top\:\! \mymat{C}_{pp}(\myvec{q}_{1i})\:\! \myvec{k}_{i} + \sum_{i=2}^{l-1}\sum_{j=i+1}^l \myvec{k}_{i}^\top \:\!\mymat{C}_{pp}(\myvec{q}_{1i} - \myvec{q}_{1j})\:\! \myvec{k}_{j} \right] \\
			& = t^2 \dpartsq{q_{1a,m}}{q_{1b,n}} \left[{\vphantom{ \sum_{j=a+1}^l}} \right.\!\myvec{k}_{1}^\top\:\! \mymat{C}_{pp}(\vec{q}_{1a})\:\! \myvec{k}_{a}\:\!\delta_{ab} +\delta_{ab}\sum_{\substack{j=2\\ j \neq a}}^{l} \myvec{k}_{a}^\top\:\! \mymat{C}_{pp}(\myvec{q}_{1a} - \myvec{q}_{1j})\:\! \myvec{k}_{j}\, +  \\
			&\quad +  (1-\delta_{ab})\,  \myvec{k}_{a}^\top \:\!\mymat{C}_{pp}(\myvec{q}_{1a} - \myvec{q}_{1b})\:\! \myvec{k}_{b}\! \left.{\vphantom{ \sum_{j=a+1}^l}} \right]  \\
			& = t^2 \dpartsq{q_{1a,m}}{q_{1b,n}} \left[{\vphantom{ \sum_{j=a+1}^l}} \right.\! \delta_{ab}\sum_{\substack{j=1\\ j \neq a}}^{l} \myvec{k}_{a}^\top\:\! \mymat{C}_{pp}(\myvec{q}_{1a} - \myvec{q}_{1j})\:\! \myvec{k}_{j}\, + \\
			&\quad + (1-\delta_{ab})\,  \myvec{k}_{a}^\top \:\!\mymat{C}_{pp}(\myvec{q}_{1a} - \myvec{q}_{1b})\:\! \myvec{k}_{b}\! \left.{\vphantom{ \sum_{j=a+1}^l}} \right]\ ,
		\end{split}
	\end{equation}
	where we removed all terms that did not depend on the variables in the partial derivatives.
	
	With these simplifications, the Hessian in zero can be calculated as
	\begin{equation}
		\begin{split}
			&\left.\frac{\partial^2\:\!Q(l)}{\partial\:\! q_{1a,m}\:\!\:\!\partial\:\! q_{1b,n}}\right|_{\myvec{q}=0} = t^2 \left.\dpartsq{q_{1a,m}}{q_{1b,n}}\right|_{\myvec{q}=0} \left[{\vphantom{ \sum_{j=a+1}^l}} \right. \\ 
			& \quad \delta_{ab}\sum_{\substack{j=1\\ j \neq a}}^{l}\, \int_\myvec{k} (\myvec{k}_{a}\cdot \myvec{k})(\myvec{k}_{j}\cdot \myvec{k})P_\psi(k)  \cos\!\left(\myvec{k}\cdot(\myvec{q}_{1a} - \myvec{q}_{1j})\right) + \\
			&\quad +(1-\delta_{ab})\!\int_\myvec{k}  (\myvec{k}_{a}\cdot \myvec{k})(\myvec{k}_{b}\cdot \myvec{k})P_\psi(k) \cos\!\left(\myvec{k}\!\cdot\!(\myvec{q}_{1a} - \myvec{q}_{1b})\right)\! \left.{\vphantom{ \sum_{j=a+1}^l}} \right] \\
			&= t^2 \left[{\vphantom{ \int_\myvec{k}}} \right.\! \delta_{ab}\, \int_\myvec{k} (\myvec{k}_{a}\cdot \myvec{k})(\myvec{k}_{a}\cdot \myvec{k})P_\psi(k) (\myvec{k})_m(\myvec{k})_n\, + \\
			&\quad + (1-\delta_{ab})\int_\myvec{k}  (\myvec{k}_{a}\cdot \myvec{k})(\myvec{k}_{b}\cdot \myvec{k})P_\psi(k) (\myvec{k})_m(\myvec{k})_n \! \left.{\vphantom{ \int_k}} \right] \\
			&= t^2 \, \int_\myvec{k}  (\myvec{k}_{a}\cdot \myvec{k})(\myvec{k}_{b}\cdot \myvec{k})P_\psi(k) (\myvec{k})_m(\myvec{k})_n\ , 
		\end{split}
	\end{equation}
	where $(\myvec{k})_m$ denotes the $m$-th component of the vector $\myvec{k}$. In performing the derivative and setting $\myvec{q}=0$, the $j$-dependence becomes very simple. This made it possible to use the overall Dirac delta condition, $\sum_{i=1}^l \myvec{k}_i=0$, to get rid of the sum. Before that, it was used that the complex phase could be rewritten as a cosine, because $\mymat{C}_{pp}$ is even. To continue evaluating, use the formulae for $m\neq n$
	\begin{align}
		&\int_k k_m^2\:\!  k_n^2 P_\psi(k) = \int_k \frac{k_m^2\:\!  k_n^2}{k^4} P_\delta(k) = \frac{1}{15} \sigma_2^2\ , \\
		&\int_k k_m^4 P_\psi(k) = \int_k \frac{k_m^4 }{k^4} P_\delta(k) = \frac{1}{5} \sigma_2^2\ , \\
		&\int_k k_m\:\!  k_n\:\! k_p^2 P_\psi(k) = 0\ .
	\end{align}
	These relations can be found by direct computation using spherical coordinates. Then they can be used to rewrite the equation from before as
	\begin{equation}
		\begin{split}
			&= t^2 \, \int_k  (\myvec{k})_m(\myvec{k})_n P_\psi(k) \left[ k_{a,x}k_{b,x}k_x^2 + k_{a,y}k_{b,y}k_y^2 \, + k_{a,z}k_{b,z}k_z^2 +\right. \\
			&\quad\left. + (k_{a,x}k_{b,y}+ k_{a,y}k_{b,x})k_xk_y + (k_{a,x}k_{b,z}+ k_{a,z}k_{b,x})k_xk_z + (k_{a,y}k_{b,z}+ k_{a,z}k_{b,y})k_yk_z {\vphantom{k_{a,x}k_{b,x}k_x^2 + k_{a,y}k_{b,y}k_y^2 + k_{a,z}k_{b,z}k_z^2 \,+}}\right] \\
			&=\frac{t^2\,\sigma_2^2}{15} \left[\delta_{mn}\left( \myvec{k}_a\cdot\myvec{k}_b + 2\cdot k_{a,m}k_{b,m}\right) + (1-\delta_{mn}) \left( k_{a,m}k_{b,n} + k_{a,n}k_{b,m} \right) {\vphantom{ \myvec{k}_a}}\right] \\
			&=\frac{t^2\,\sigma_2^2}{15} \left[ \left( k_{a,m}k_{b,n} + k_{a,n}k_{b,m} \right) + \delta_{mn}\, \vec{k}_a\cdot\vec{k}_b \right]\ .
		\end{split}
	\end{equation}
	Writing the $m$, $n$ indices as blocks of the Hessian matrix gives the final expression
	\begin{equation}
		\left.\frac{\partial^2\:\!Q(l)}{\partial\:\! q_{1a,m}\:\!\:\!\partial\:\! q_{1b,n}}\right|_{\mathbf{q}=0} = \frac{t^2\,\sigma_2^2}{15} \left( \vec{k}_a \otimes \vec{k}_b + \vec{k}_b \otimes \vec{k}_a +\,(\vec{k}_a \cdot \vec{k}_b)\,\idmat_3 \right)_{m,n} \ .
	\end{equation}
	
	\subsection{Isolated Critical Point}
	This Section will address the last point in the list of prerequisites. Namely that the critical point in $\myvec{q}=0$ is isolated, i.e., there is no other critical point infinitesimally close to zero. This is in general a hard problem as the interest is in the typical behaviour in relation to the $\myvec{k}$ parameters in the sense that there might be a very particular configuration that shows degeneracies, but all other configurations are fine. There is a mathematically precise way of analyzing this based on \cite{post78}, a typicality argument based on René Thom. A formal argument for the general $Q(l)$ is beyond the scope of this work however.
	
	This Section will focus on the specific case of the bispectrum in a non-degenerate $\myvec{k}$ configuration: $k_{2z}\neq 0$, $k_{3x}\neq 0$, $k_{2z}+k_{3z}\neq 0$. This is the main interest of this work and will turn out to be easier to compute as the Hessian has corank $1$.
	
	Looking at the quadratic form $Q_f$ in equations~(\ref{eq:bispec:quadform1})-(\ref{eq:bispec:quadform2}), the first impulse is to expand it for small $\myvec{q}$. This however yields an approximation of the type
	\begin{equation}
		Q_f \approx \frac{1}{2}\myvec{q}^t \mymat{H} \myvec{q}\ , 
	\end{equation}
	with $\myvec{q}$ the $6$-dimensional variable vector and $\mymat{H}$ the Hessian matrix of $Q_f$ in $\myvec{q}=0$. As noted in the main part, $\operatorname{rank}(\mymat{H})=5$, thus there is a straight line of critical points in this approximation.
	
	This can be used as an advantage: calculate the 1-dimensional kernel of $\mymat{H}$ and then check that the derivative does not vanish on this line any longer for a higher-order approximation of $Q_f$.
	
	Doing this, it is straightforward to see that 
	\begin{equation}
		\operatorname{ker}(\mymat{H}) = \operatorname{span}\!\left(\myvec{q}_\text{ker}\right) \coloneqq \operatorname{span}\!\left( \left( -\frac{k_{3x}}{k_{2z}}, 0, -\frac{k_{3z}}{k_{2z}} , 0 , 0 , 1 \right)^t \right)
	\end{equation}
	is the explicit description for the kernel. Using an approximation up to fourth order in $\myvec{q}$ of $Q_f$, calculate the derivative and look at
	\begin{equation}
		\begin{split}
			\left.\frac{\partial Q_f}{\partial \myvec{q}} \right|_{\myvec{q} =\epsilon \myvec{q}_\text{ker}} &= \frac{\sigma_3^2 t^2 \epsilon^3}{210 k_{2z}^2} \left(-k_{2z} k_{3x} (k_{2z}^2 + 2 k_{3x}^2 + 6 k_{2z} k_{3z} +6 k_{3z}^2), 0, -k_{2z} (k_{2z} +\right.\\
			&\quad + 2 k_{3z}) (3 k_{3x}^2 + 5 k_{3z} (k_{2z} + k_{3z})), k_{3x} (3 k_{2z} k_{3x}^2 + 4 (k_{2z}^2 +k_{3x}^2) k_{3z} + \\
			&\quad +9 k_{2z} k_{3z}^2 + 4 k_{3z}^3), 0, k_{3x}^4 + 6 k_{3x}^2 k_{3z}^2 + 5 k_{3z}^4 + 2 k_{2z}^2 (k_{3x}^2 + 5 k_{3z}^2) +\\
			&\quad \left.+ 3 k_{2z} (3 k_{3x}^2 k_{3z} + 5 k_{3z}^3) \right)
		\end{split}
	\end{equation}
	where $\epsilon$ is used to parameterize the one-dimensional space $\operatorname{ker}(\mymat{H})$. This obviously gives a critical point at $\epsilon=0$, i.e., $\myvec{q}=0$. Note the conditions on non-degeneracy considered here. We also assume $\sigma_3^2\neq 0$; but even if zero, a higher-order approximation could then be considered. To obtain a critical point for $\epsilon\neq 0$, the parenthesis in 4 components would have to compute to zero. This can be tested, but is only possible at the same time for $k_{2z}=k_{3x}=k_{3z}=0$, which we exclude here. This can also be checked to even higher order with Mathematica, but the same pattern exists. Thus the critical point is shown to be isolated.
	
	For the parallel configuration $k_{3x}=0$, the corank of the Hessian is 3, so the problem becomes harder to compute. For higher polyspectra this is the same case. Then either a typicality argument needs to be leveraged or the following physical argument: a non-isolated critical point means that there is at least a line of critical points leading away from the original point. Close up, this looks like a one-dimensional subspace. This would mean that a direction in $\myvec{q}$ space is preferred, in contradiction to isotropy. Alternatively, critical points could form a sphere around zero, in line with isotropy. But this can be easily disproven with the same technique as before. It is easy to check that for example along the first coordinate axis, there is only a critical point in zero.
	
	\section{Details on the Asymptotic Bispectrum} \label{sec:app:bispec}
	In the main Sect.~\ref{sec:methods:higherorder} it was already detailed how higher orders could be added systematically. To understand this, we will firstly look at monomials in front of the exponential. The prefactor essentially scales with $k^2$ and with the formulas for $n\in\N_0$ 
	\begin{equation}
		\int \mathrm{d}q \,q^{2n} \expr{-k^2 q^2} \sim \frac{1}{k^{1+2n}}\ , \quad \int \mathrm{d}q \,q^{2n} \expr{-k^2 q^4} \sim \frac{1}{k^\frac{1+2n}{2}} \ , \quad \int \mathrm{d}q \,q^{2n} \expr{-k^2 q^2+\iu k q} \sim \frac{1}{k^{1+2n}}
	\end{equation}
	it is clear to see which terms are relevant for the next orders. The $k^{-11/2}$ order is generated by a $1$ in front of the exponential. The next order $k^{-13/2}$ is generated by $q_{6}^2$ monomials in front of the full exponential, while there is no dependence on $q_i$, $i\in\{1,\dots,5\}$, in the terms in front. The last order considered $k^{-15/2}$ is generated by either $q_i^2$, for one $i\in\{1,\dots,5\}$, or $q_{6}^4$ monomials. 
	
	The first subsequent transformation contains the identity transformation and order 3 monomials. The second subsequent transformation contains the identity transformation and order 5 monomials. So these are the only subsequent transformations that need to be taken into account, as the determinant of their Jacobians are the only one contributing to degree 2 and 4 in front of the exponential.
	
	The final transformation based on the residual polynomial $g(q_{13z})$ is more complicated to take into account. Computationally, higher orders in $q_{13z}$ are considered by performing a Taylor expansion. This produces terms in front of the exponential of degree $6$ and $8$. This is however fine, as they come with a factor of $k^2$, thus still contribute to the next-order approximation.
	
	In the end, the subsequent transformations have an effect on the complex phase as well. Here the same procedure is applied, expanding the exponentials of these higher-order contributions. 
	
	Combined, this leads to these terms in front of the exponential of equation~(\ref{eq:bispec:general}), given by
	\begin{equation}
		\begin{split}
			J &\approx\left(1+j_1^{(2)}+j_1^{(4)}\right)\left(1+j_2^{(4)}\right)\left(1+t_g^{(6)}+t_g^{(8)}\right)\left(1+t_{1,\phi}^{(3)}\right) \\
			&\approx 1+j_1^{(2)}+j_1^{(4)}+j_2^{(4)}+t_g^{(6)}+t_g^{(8)}+t_{1,\phi}^{(3)}+j_1^{(2)}t_g^{(6)}\ ,
		\end{split}
	\end{equation}
	where $j_1$ and $j_2$ are the Jacobians of given order of the first and second subsequent transformations respectively. $t_f$ describes the terms obtained by expanding the residual polynomial $g$ and $t_{1,\phi}$ the terms obtained by expanding the higher-order complex phase parts. 
	
	The results of the calculations are the following. For the next-order (NO) contribution, we can split this into the original result $\bispec_{k\rightarrow\infty}=\bispec^{(0)}_{k\rightarrow\infty}$ from Eq.~(\ref{eq:bispec:splittingresult}), and higher orders $\bispec^{(1)}_{k\rightarrow\infty}$ and $\bispec^{(2)}_{k\rightarrow\infty}$ given by
	\begin{equation}
		\bispec^{(1)}+\bispec^{(2)} = \frac{c_0^\text{NO} \cdot \expr{-\frac{15}{4}\left(t\,\sigma_2\right)^{-2}}\, \sqrt{\abs{k_{2z}}}\, \left(f^{(1)}+f^{(2)}\right)}{t^{11}\,\abs{\sigma_2}^{13} \sqrt{t\abs{\sigma_3}^{11}}\left(c_4(\mathbf{k})\right)^\frac{11}{4}\:k_{3x}^6\,k_{2z}^\frac{9}{2}}
	\end{equation}
	for large $k$, with the numerical coefficient
	\begin{equation}
		c_0^\text{NO} = \frac{5\,\sqrt[\leftroot{2}4]{5}\: \pi^{\frac{5}{2}}}{14784\sqrt{3}\,\sqrt[\leftroot{2}4]{7\cdot 8}}\approx 0.00187 
	\end{equation}
	and the functions $f^{(1)}$, $f^{(2)}$ encoding the complex $\myvec{k}$ dependence. They are given by
	\begin{align*}
		&f^{(1)}(\myvec{k}_2,\myvec{k}_3,\sigma,t) =\\
		&\, - 61600\, t^3\sigma_2^4\sigma_3^2k_{2z}k_{3x}^2\Gamma\!\left(\frac{7}{4}\right) \left(k_{3x}^4 + 6 k_{3x}^2 k_{3z}^2 + 5 k_{3z}^4 + k_{2z}^2 (k_{3x}^2 + 5 k_{3z}^2) + 2 k_{2z} \cdot \right.\\
		&\,\left. \cdot (3 k_{3x}^2 k_{3z} + 5 k_{3z}^3)\right) \left[k_{2z}^4\left(1215\sigma_3^4 (k_{3x}^4 + 2 k_{3x}^2 k_{3z}^2 - 15 k_{3z}^4)+ 2 t^2\sigma_2^2 (27 \sigma_3^4(15 k_{3x}^4 +\right.\right.\\
		&\, \left.+ 14 k_{3x}^2 k_{3z}^2 +495 k_{3z}^4) -140 \sigma_2^2\sigma_4^2 k_{3x}^2 (k_{3x}^2 + 7 k_{3z}^2))\right)+\left(k_{3x}^2+k_{3z}^2\right)^2 \left(1215 (k_{3x}^4 + \right. \\
		&\, +2 k_{3x}^2 k_{3z}^2 - 15 k_{3z}^4) \sigma_3^4 + 2 t^2 \sigma_2^2 (27 (15 k_{3x}^4 + 14 k_{3x}^2 k_{3z}^2 + 495 k_{3z}^4) \sigma_3^4 - 140 k_{3x}^2 (k_{3x}^2 + \\
		&\, \left.+7 k_{3z}^2)\sigma_2^2 \sigma_4^2)\right) + 12 k_{2z}^3 k_{3z} \left(405 (k_{3x}^4 - 6 k_{3x}^2 k_{3z}^2 - 15 k_{3z}^4) \sigma_3^4 + 2 \sigma_2^2 (9 (11 k_{3x}^4 + \right. \\
		&\, \left.+258 k_{3x}^2 k_{3z}^2 + 495 k_{3z}^4) \sigma_3^4 - 35 k_{3x}^2 (3 k_{3x}^2 + 7 k_{3z}^2) \sigma_2^2 \sigma_4^2) t^2\right) +12 k_{2z} k_{3z} \left(k_{3x}^2 + k_{3z}^2\right) \cdot \\
		&\,\cdot \left(405 (k_{3x}^4 -6 k_{3x}^2 k_{3z}^2 - 15 k_{3z}^4) \sigma_3^4 + 2 \sigma_2^2 (9 (11 k_{3x}^4 + 258 k_{3x}^2 k_{3z}^2 + 495 k_{3z}^4) \sigma_3^4 \right. -\\
		&\, \left. - 35 k_{3x}^2 (3 k_{3x}^2 + 7 k_{3z}^2) \sigma_2^2 \sigma_4^2) t^2\right) + \left.2 k_{2z}^2 \right(\!1215 \sigma_3^4 (k_{3x}^6 - 3 k_{3x}^4 k_{3z}^2 - 41 k_{3x}^2 k_{3z}^4 - 45 k_{3z}^6) -\\
		&\,- 2 \sigma_2^2 (27 (17 k_{3x}^6 -291 k_{3x}^4 k_{3z}^2 - 1513 k_{3x}^2 k_{3z}^4 - 1485 k_{3z}^6) \sigma_3^4 + \\
		&\,\left.\left. + 56 k_{3x}^2 (3 k_{3x}^4 + 30 k_{3x}^2 k_{3z}^2 + 35 k_{3z}^4) \sigma_2^2 \sigma_4^2) t^2\right)\right]
	\end{align*}
	and
	\begin{align*}
		&f^{(2)}(\myvec{k}_2,\myvec{k}_3,\sigma,t) = 5\sqrt{70}\,\Gamma\!\left(\frac{9}{4}\right) \abs{\sigma_3}\cdot\\
		&\, \cdot \sqrt{(k_{3x}^4 + 6 k_{3x}^2 k_{3z}^2 + 5 k_{3z}^4 + k_{2z}^2 (k_{3x}^2 + 5 k_{3z}^2) + 2 k_{2z} (3 k_{3x}^2 k_{3z} + 5 k_{3z}^3)) \sigma_3^2}\, \cdot \\
		&\,\cdot(52 k_{2z}^5 k_{3z} (1804275 (k_{3x}^2 + 5 k_{3z}^2) (19 k_{3x}^4 + 28 k_{3x}^2 k_{3z}^2 - 15 k_{3z}^4) \sigma_3^6 - 160380 (k_{3x}^2 + 5 k_{3z}^2)\cdot \\
		&\, \cdot (593 k_{3x}^4 + 668 k_{3x}^2 k_{3z}^2 - 765 k_{3z}^4) \sigma_2^2 \sigma_3^6 t^2 - 2 \sigma_2^4 (-297 (19087 k_{3x}^6 - 80489 k_{3x}^4 k_{3z}^2 + \\
		&\, + 502785 k_{3x}^2 k_{3z}^4 - 108135 k_{3z}^6) \sigma_3^6 + 308 k_{3x}^2 (24131 k_{3x}^4 + 109190 k_{3x}^2 k_{3z}^2 - 37125 \cdot \\
		&\, \cdot k_{3z}^4) \sigma_2^2 \sigma_3^2 \sigma_4^2 +329280 k_{3x}^4 (k_{3x}^2 + 3 k_{3z}^2) \sigma_2^4 \sigma_5^2) t^4) + 2 k_{2z} k_{3z} (1804275 (k_{3x}^2 + k_{3z}^2)^2 \cdot \\
		&\, \cdot (k_{3x}^2 + 5 k_{3z}^2) (19 k_{3x}^4 + 28 k_{3x}^2 k_{3z}^2 - 15 k_{3z}^4) \sigma_3^6 - 160380 (k_{3x}^2 + k_{3z}^2) (k_{3x}^2 + 5 k_{3z}^2) \cdot \\
		&\, \cdot (1025 k_{3x}^6 + 1837 k_{3x}^4 k_{3z}^2 -337 k_{3x}^2 k_{3z}^4 - 765 k_{3z}^6) \sigma_2^2 \sigma_3^6 t^2 - 2 \sigma_2^4 (297 (127793 k_{3x}^{10} + \\
		&\, + 1502475 k_{3x}^8 k_{3z}^2 + 3915906 k_{3x}^6 k_{3z}^4 + 2423054 k_{3x}^4 k_{3z}^6 -286515 k_{3x}^2 k_{3z}^8 + 108135 \cdot \\
		&\,\cdot k_{3z}^{10}) \sigma_3^6 + 308 k_{3x}^2 (k_{3x}^2 + k_{3z}^2) (40331 k_{3x}^6 + 252121 k_{3x}^4 k_{3z}^2 + 261065 k_{3x}^2 k_{3z}^4 - 37125 \cdot \\
		&\,\cdot k_{3z}^6) \sigma_2^2 \sigma_3^2 \sigma_4^2 + 329280 k_{3x}^4 (k_{3x}^2 + k_{3z}^2)^2 (k_{3x}^2 + 3 k_{3z}^2) \sigma_2^4 \sigma_5^2) t^4) + k_{2z}^6 (1804275 (3 k_{3x}^2 -\\
		&\,- k_{3z}^2) (k_{3x}^2 + 5 k_{3z}^2)^2 \sigma_3^6 - 160380 (97 k_{3x}^2 -51 k_{3z}^2) (k_{3x}^2 + 5 k_{3z}^2)^2 \sigma_2^2 \sigma_3^6 t^2 + 2 \sigma_2^4 (297 \cdot \\
		&\,\cdot (12863 k_{3x}^6 + 28385 k_{3x}^4 k_{3z}^2 + 222741 k_{3x}^2 k_{3z}^4 - 36045 k_{3z}^6) \sigma_3^6 + 616 k_{3x}^2 (43 k_{3x}^4 -\\
		&\,- 15260 k_{3x}^2 k_{3z}^2 + 7425 k_{3z}^4) \sigma_2^2 \sigma_3^2 \sigma_4^2 - 54880 k_{3x}^4 (k_{3x}^2 +9 k_{3z}^2) \sigma_2^4 \sigma_5^2) t^4) + (k_{3x}^2 + k_{3z}^2)\cdot\\
		&\,\cdot (1804275 (3 k_{3x}^2 - k_{3z}^2) (k_{3x}^4 +6 k_{3x}^2 k_{3z}^2 + 5 k_{3z}^4)^2 \sigma_3^6 - 160380 (k_{3x}^2 + k_{3z}^2) (k_{3x}^2 + 5 k_{3z}^2)^2 \cdot \\
		&\, \cdot (169 k_{3x}^4 + 22 k_{3x}^2 k_{3z}^2 - 51 k_{3z}^4) \sigma_2^2 \sigma_3^6 t^2 - 2 \sigma_2^4 (297 (1537 k_{3x}^{10} +254049 k_{3x}^8 k_{3z}^2 + \\
		&\,+ 1248426 k_{3x}^6 k_{3z}^4 + 1362178 k_{3x}^4 k_{3z}^6 - 150651 k_{3x}^2 k_{3z}^8 + 36045 k_{3z}^{10}) \sigma_3^6 + 616 k_{3x}^2 \cdot \\
		&\,\cdot (k_{3x}^2 + k_{3z}^2) (1757 k_{3x}^6 +36817 k_{3x}^4 k_{3z}^2 + 70835 k_{3x}^2 k_{3z}^4 - 7425 k_{3z}^6) \sigma_2^2 \sigma_3^2 \sigma_4^2 + 54880 \cdot \\
		&\, \cdot k_{3x}^4 (k_{3x}^2 + k_{3z}^2)^2 (k_{3x}^2 + 9 k_{3z}^2) \sigma_2^4 \sigma_5^2) t^4) + k_{2z}^4 (1804275 (9 k_{3x}^8 + 228 k_{3x}^6 k_{3z}^2 + 742 k_{3x}^4 \cdot \\
		&\, \cdot k_{3z}^4 + 340 k_{3x}^2 k_{3z}^6 - 375 k_{3z}^8) \sigma_3^6 - 160380(363 k_{3x}^8 + 7500 k_{3x}^6 k_{3z}^2 + 20402 k_{3x}^4 k_{3z}^4 - \\
		&\, - 1060 k_{3x}^2 k_{3z}^6 - 19125 k_{3z}^8) \sigma_2^2 \sigma_3^6 t^2 - 2 \sigma_2^4 (-297 (1213 k_{3x}^8 - 703436 k_{3x}^6 k_{3z}^2 -1836386 \cdot \\
		&\, \cdot k_{3x}^4 k_{3z}^4 + 1909476 k_{3x}^2 k_{3z}^6 - 540675 k_{3z}^8) \sigma_3^6 + 616 k_{3x}^2 (7529 k_{3x}^6 + 168187 k_{3x}^4 k_{3z}^2 + \\
		&\,+ 307255 k_{3x}^2 k_{3z}^4 - 81675 k_{3z}^6) \sigma_2^2 \sigma_3^2 \sigma_4^2 + 188160 k_{3x}^4 (k_{3x}^4 + 14 k_{3x}^2 k_{3z}^2 + 21 k_{3z}^4) \sigma_2^4 \sigma_5^2) \cdot \\
		&\, \cdot t^4) + k_{2z}^2 (1804275 (k_{3x}^2 + k_{3z}^2) (9 k_{3x}^8 + 228 k_{3x}^6 k_{3z}^2 + 742 k_{3x}^4 k_{3z}^4 + 340 k_{3x}^2 k_{3z}^6 - 375 \cdot \\
		&\, \cdot k_{3z}^8) \sigma_3^6 - 160380 (435 k_{3x}^{10} + 11223 k_{3x}^8 k_{3z}^2 +37934 k_{3x}^6 k_{3z}^4 + 24622 k_{3x}^4 k_{3z}^6 - 23185 \cdot\\
		&\,\cdot k_{3x}^2 k_{3z}^8 - 19125 k_{3z}^{10}) \sigma_2^2 \sigma_3^6 t^2 - 2 \sigma_2^24 (297 (59267 k_{3x}^{10} + 1908943 k_{3x}^8 k_{3z}^2 + 8573422 k_{3x}^6\cdot \\
		&\, \cdot k_{3z}^4 + 7486910 k_{3x}^4 k_{3z}^6 - 1368801 k_{3x}^2 k_{3z}^8 + 540675 k_{3z}^{10}) \sigma_3^6 + 616 k_{3x}^2 (10049 k_{3x}^8 + \\
		&\,+ 217116 k_{3x}^6 k_{3z}^2 + 657242 k_{3x}^4 k_{3z}^4 + 414580 k_{3x}^2 k_{3z}^6 - 81675 k_{3z}^8) \sigma_2^2 \sigma_3^2 \sigma_4^2 + 188160 k_{3x}^4 \cdot \\
		&\, \cdot (k_{3x}^2 + k_{3z}^2) (k_{3x}^4 +14 k_{3x}^2 k_{3z}^2 + 21 k_{3z}^4) \sigma_2^4 \sigma_5^2) t^4) + 4 k_{2z}^3 k_{3z} (1804275 (19 k_{3x}^8 + 160 k_{3x}^6 \cdot \\
		&\,\cdot k_{3z}^2 + 290 k_{3x}^4 k_{3z}^4 + 40 k_{3x}^2 k_{3z}^6 - 125 k_{3z}^8) \sigma_3^6 - 160380 (809 k_{3x}^8 + 5792 k_{3x}^6 k_{3z}^2 + 7270 \cdot \\
		&\,\cdot k_{3x}^4 k_{3z}^4 - 4360 k_{3x}^2 k_{3z}^6 - 6375 k_{3z}^8) \sigma_2^2 \sigma_3^6 t^2 - 2 \sigma_2^4 (297 (101617 k_{3x}^8 + 1023456 k_{3x}^6 k_{3z}^2 + \\
		&\, + 1374838 k_{3x}^4 k_{3z}^4 - 509256 k_{3x}^2 k_{3z}^6 + 180225 k_{3z}^8) \sigma_3^6 + 308 k_{3x}^2 (35445 k_{3x}^6 + 229587 k_{3x}^4\cdot \\
		&\,\cdot k_{3z}^2 + 229975 k_{3x}^2 k_{3z}^4 - 51975 k_{3z}^6) \sigma_2^2 \sigma_3^2 \sigma_4^2 + 19600 k_{3x}^4 (15 k_{3x}^4 + 70 k_{3x}^2 k_{3z}^2 + 63 k_{3z}^4) \cdot \\
		&\,\cdot \sigma_2^4 \sigma_5^2) t^4)) 
	\end{align*}
	The polynomial $c_4(\mathbf{k})$ of degree and order $4$ is the same as for the leading order approximation.
	
	\section{Details on the Asymptotic Bispectrum with Degeneracy} \label{sec:app:bispecdegen}
	This Section gives details on the calculation of the asymptotic bispectrum in a degenerate configuration of $\myvec{k}$ vectors. Only the parallel situation of $\myvec{k}_2=k_{2z}\hat{\myvec{e}}_z$ and $\myvec{k}_3=k_{3z}\hat{\myvec{e}}_z$ is considered with $k_{2z}\neq 0$, $k_{3z}\neq 0$, and $k_{2z}+k_{3z}\neq 0$. 
	
	When applying the proof of the splitting lemma, start with a coordinate transform that gives a diagonal Hessian. The transformation due to Sylvester's law of inertia applied to $\frac{1}{2}\mymat{H}^{\parallel}$ is modified like before to achieve a determinant of $1$ while keeping the entries depending on the $\myvec{k}_i$ components ``balanced''. This yields
	\begin{equation}
		\mymat{T}^{\parallel} = \begin{pmatrix} 1&0&0&-\frac{k_{3z}}{k_{2z}}&0&0\\
			0&1&0&0&-\frac{k_{3z}}{k_{2z}}&0\\
			0&0&1&0&0&-\frac{k_{3z}}{k_{2z}}\\
			0&0&0&1&0&0\\
			0&0&0&0&1&0\\
			0&0&0&0&0&1 \end{pmatrix}\ ,
	\end{equation}
	which is coincidentally the result from the non-degenerate case with $k_{3x}=0$. Note that this is not true for the original, ``unbalanced'' transformation. The result after transformation is then
	\begin{equation}
		\frac{1}{2} \left(\mymat{T}^{\parallel}\right)^\top \mymat{H}^{\parallel} \mymat{T}^{\parallel} = \frac{t^2\sigma_2^2}{30} \operatorname{diag}\left(k_{2z}^2,k_{2z}^2,3k_{2z}^2,0,0,0\right)\ .
	\end{equation}
	Applying the subsequent transformations yields a residual part $g^{\parallel}(q_{13x},q_{13y},q_{13z})$ where notationally the original and the transformed variables are not distinguished. This residual part has order $4$, but cannot be simplified further by the splitting lemma alone. 
	
	As the main interest here is not the precise normal form of the residual part, but the asymptotic description of the whole integrand of the bispectrum, this should not be an issue. Then either use that $g^{\parallel}$ should be finitely determined because zero is an isolated critical point or, as this has not been proven rigorously, use that $a_1(q)$ and $a_2(q)$ are absolutely convergent for exponential cut-offs \cite{konr20}, and $g^{\parallel}$ should at least be a convergent power series. Thus the high-order monomials should be suppressed in the limit $q\rightarrow 0$. For the leading order term, truncate $g^{\parallel}$ at degree four to get
	\begin{equation}
		\begin{split}
			g^{\parallel}&\approx \frac{t^2 \sigma_3^2}{280} \frac{k_{3z}^2(k_{2z}+k_{3z})^2}{k_{2z}^2}\myvec{q}_{13}^2 \left(\myvec{q}_{13}^2+4q_{13z}^2\right)
		\end{split}
	\end{equation}
	for $q\rightarrow 0$. It has been carefully checked in Mathematica \cite{mathematica}, that no further subsequent transformations of the splitting lemma influence this description. Additionally, applying only the first transformation $\mymat{T}^{\parallel}$ to the phase part yields
	\begin{equation}
		\iu \left(k_{2z}q_{12z}+k_{3z}q_{13z}\right) \quad \mapsto \quad \iu\, k_{2z}q_{12z}\ .
	\end{equation}
	This description can then be plugged into the formula of the free bispectrum, equation~(\ref{eq:bispec:general}). The quadratic and quartic parts in the exponential separate in the integration. Computing the integral over the Gaussians is simple and yields
	\begin{equation}
		\frac{30\sqrt{10}\,\pi^{\frac{3}{2}}\cdot\expr{-\frac{5}{2}\left(t\,\sigma_2\right)^{-2}}}{\left(t\abs{\sigma_2}\right)^3 \abs{k_{2z}}^{3}}\ .
	\end{equation}
	The quartic part is slighly more complicated because it is not decoupled. However, using spherical coordinates $|\myvec{q}_{13}|\eqqcolon q_{13}$
	\begin{equation}
		(q_{13x},q_{13y},q_{13z}) \quad \mapsto \quad (q_{13},\varphi_{13},\theta_{13})\ ,
	\end{equation}
	the residual part becomes
	\begin{equation}
		g^{\parallel}\approx \frac{t^2 \sigma_3^2}{280} \frac{k_{3z}^2(k_{2z}+k_{3z})^2}{k_{2z}^2}\, q_{13}^4\left(3+2\cos\!\left(2\,\theta_{13}\right)\right)
	\end{equation}
	for small $q_{13}$. For the radial component, use the formula
	\begin{equation}
		\int_0^\infty \mathrm{d} r \, r^2\, \expr{-a\cdot r^4} = \frac{1}{4} \Gamma\!\left(\frac{3}{4}\right) a^{-\frac{3}{4}}\ ,
	\end{equation}
	with $a\in\R^+$. Then the integral can be partially performed over $q_{13}$ and $\varphi_{13}$, which gives
	\begin{equation}
		\begin{split}&\int_{0}^\pi \mathrm{d}\theta_{13}\int_0^{2\pi}\mathrm{d}\varphi_{13}\int_0^\infty\mathrm{d} q_{13}\, \sin(\theta_{13})\,q_{13}^2 \,\expr{-g^{\parallel}} \\
			&= -\pi\left(\frac{35}{2}\right)^{\frac{3}{4}} \Gamma\!\left(-\frac{1}{4}\right) \left(\frac{t^2k_{3z}^2(k_{2z}+k_{3z})^2\sigma_3^2}{k_{2z}^2}\right)^{\frac{3}{4}}\cdot \int_{0}^\pi \mathrm{d}\theta_{13}\, \frac{\sin(\theta_{13})}{(3+2\cos(2\,\theta_{13}))^{\frac{3}{4}}}\ .\end{split}
	\end{equation}
	The last integral over $\theta_{13}$ can be transformed using $u=\cos(\theta_{13})$ as
	\begin{equation}
		\int_{0}^\pi \mathrm{d}\theta_{13}\, \frac{\sin(\theta_{13})}{(3+2\cos(2\,\theta_{13}))^{\frac{3}{4}}} = \int_{-1}^1 \mathrm{d} u\, \frac{1}{(1+4u^2)^{\frac{3}{4}}}= 2 \cdot\operatorname{_2F_1}\!\left(\frac{1}{2},\frac{3}{4} ;\frac{3}{2},-4 \right)\ ,
	\end{equation}
	using the hypergeometric $\operatorname{_2F_1}$ function, for which efficient numerical implementations exist.
	
	The result for the bispectrum is then simply the product of the quadratic and the quartic result. Note that it was never guaranteed that the truncation at degree $4$ would work, the integral could have diverged and more terms of higher orders would have had to be taken into account.

	\section{Details on the Numerical Implementation} \label{sec:app:numerics}
	\subsection{Details on Cosmological Model}
	The precise numerical values of some cosmological constants were not given previously. All measured constants are taken from the Planck collaboration \cite{agha20}.
	
	The Bardeen power spectrum for cold dark matter was used for the initial density perturbation power spectrum \cite{bard86}. Its transfer function is given by
	\begin{equation}
		T_\text{Bardeen}(k) = \frac{\ln(1+2.34q)}{2.34q} \left[ 1+3.89q + (16.1q)^2+ (5.46)^3 + (6.71q)^4 \right]^{-\frac{1}{4}}\ ,
	\end{equation}
	where $q$ is given by 
	\begin{equation}
		q = \frac{k}{h\, \operatorname{Mpc}^{-1}}\, \Omega_{X0}\, \exp\,\left( \Omega_{b0}\left( 1+ \frac{\sqrt{2h}}{\Omega_{X0}} \right) \right)\ .
	\end{equation}
	The two constants depending on cosmology occurring are $\Omega_{X0}$ and $\Omega_{b0}$, the dark matter and baryon density parameter, respectively. The values used were
	\begin{equation}
		\Omega_{X0} = 0.26\ , \quad \Omega_{b0} = 0.04\ .
	\end{equation}
	These approximate the measured values by \cite{agha20} of
	\begin{align}
		\Omega_{X0}^\text{Planck} &= (0.315 \pm 0.007) - (0.0486 \pm 0.0010) = 0.266 \pm 0.007\ , \\
		\Omega_{b0}^\text{Planck} &= 0.0486 \pm 0.0010\ .
	\end{align}
	The dimensionless Hubble constant $h$ was set to $h = 0.7$. The initial density perturbation power spectrum can then be computed with the transfer function via
	\begin{equation}
		P_\delta^\text{Bardeen}(k) = A(\sigma_8)\, T^2_\text{Bardeen}(k)\, k^{n_s}\ ,
	\end{equation}
	where $A$ is an amplitude determined $\sigma_8$ and $n_s$ is the spectral index. The parameter $\sigma_8$ describes the variance with a smoothing at $R=8\, h^{-1}\operatorname{Mpc}$. The formula for general $R$ is
	\begin{equation}
		\sigma_R^2 = \frac{1}{2\pi^2} \int \mathrm{d} k\, k^2 P_\delta(k) \hat{W}_R^2(k)\ ,
	\end{equation}
	where $\hat{W}_R(K)$ describes the Fourier transform of a filter function $W_R$ in real space. The value published by the Planck collaboration is $0.811 \pm 0.006$, the value implemented in the code was $0.8$. The spectral index used was $n_s=1.0$, while the measured one is at $0.965 \pm 0.004$. 
	
	Finally, the initial power spectra can be evolved linearly via
	\begin{equation}
		P^\text{lin}(k,t) = D^2_+\, P^\mathrm{(i)}(k)= t^2\, P^\mathrm{(i)}(k)\ .
	\end{equation}
	The initial point used in this work is the CMB release, put by the Planck collaboration at a redshift of  $1089.80 \pm 0.21$. Thus, an approximate value for the initial scale factor $a^\mathrm{(i)}=10^{-3}$ is used. In general, the time coordinate now, i.e., scale factor $a=1$, is given by $t= 901.067$. 
	
	Note that the numerics and the asymptotics described here are valid for a general class of initial power spectra and cosmological parameters. For example, the precise values of $\sigma_2^2$ will depend on these choices, but the overall dependence on $\sigma_2^2$ will remain universal.
	
	\subsection{Details on Numerical Bispectrum}
	Here the parameters used to generate the data for Fig.~\ref{fig:bispec:result} are reported. The Bardeen power spectrum was used as the initial power spectrum, as detailed in the previous Section. A logarithmic grid was used for the $k$ values, 200 points ranging from $5\cdot 10^{-5} \, h\operatorname{Mpc}^{-1}$ to $10^{4} \, h\operatorname{Mpc}^{-1}$. The Monte Carlo VEGAS algorithm was used for the numerical integration, with $10^7$ function evaluations for every result. There was no need to adjust any other parameter in the integration bound setting from their implemented defaults. 
	
	For the asymptotics the explicit moments of the initial power spectrum are needed. Here we give them for two cut-off values $k_s$ to the relevant degree, as both have been used later in Figs.~\ref{fig:bispec:angulardep} and \ref{fig:bispec:timedep}. 
	\begin{itemize}
		\item $k_s=10^2 \, h\operatorname{Mpc}^{-1}$:
		\begin{equation}
			\begin{split}
				\sigma_{2}^2 &= 3.25\cdot 10^{-4}\\
				\sigma_{3}^2 &= 0.491\, h^2\operatorname{Mpc}^{-2}\\
				\sigma_{4}^2 &= 9160\, h^4\operatorname{Mpc}^{-4}\\
				\sigma_{5}^2 &= 5.11\cdot 10^8\, h^6\operatorname{Mpc}^{-6}
			\end{split}
		\end{equation}
		\item $k_s=10^4 \, h\operatorname{Mpc}^{-1}$:
		\begin{equation}
			\begin{split}
				\sigma_{2}^2 &= 1.65\cdot 10^{-3}\\
				\sigma_{3}^2 &= 1.35\cdot 10^4\, h^2\operatorname{Mpc}^{-2}
			\end{split}
		\end{equation}
	\end{itemize}
	Please also note that the full implementation is available at \href{https://lin0.thphys.uni-heidelberg.de:4443/kft-bispec/bispec-asymp-paper}{GitLab ITP}.

	\bibliographystyle{JHEP}
	\bibliography{literature.bib}
	
\end{document}